\documentclass[fleqn,usenatbib]{mnras}

\usepackage{newtxtext,newtxmath}

\usepackage[T1]{fontenc}

\DeclareRobustCommand{\VAN}[3]{#2}
\let\VANthebibliography\thebibliography
\def\thebibliography{\DeclareRobustCommand{\VAN}[3]{##3}\VANthebibliography}

\usepackage{graphicx}	
\usepackage{amsmath}	
\usepackage{multirow}

\defcitealias{Boughelilba_2023}{BR23}
\usepackage{subcaption}

\title[FR0 radio galaxy jets]{FR0 radio galaxy jets -- I. linking jet dynamics and high-energy emission in LEDA~55267 and LEDA~58287}

\author[A. F. S. Cardoso et al.]{
A. F. S. Cardoso,$^{1}$\thanks{E-mail: a.cardoso.fis@gmail.com}
R. C. Anjos,$^{1,2,3,4,5}$
\\
$^{1}$PPGCosmo, CCE, Universidade Federal do Espírito Santo, Av. Fernando Ferrari, 540, 29075--910 Vitória, ES, Brazil\\
$^{2}$Departamento de Engenharias e Exatas, Universidade Federal do Paraná (UFPR), Pioneiro, 2153, 85950-000 Palotina, PR, Brazil\\
$^{3}$Programa de Pós-Graduação em Física e Astronomia, Universidade Tecnológica Federal do Paraná, Curitiba, PR 80230-901, Brazil\\
$^{4}$Programa de pós-graduação em Física, Universidade Estadual de Londrina (UEL), Rodovia Celso Garcia Cid Km 380, 86057-970 Londrina, PR, Brazil\\
$^{5}$Programa de Pós-Graduação em Física Aplicada, Universidade Federal da Integração Latino-Americana, 85867-670, Foz do Igua\c{c}u, PR, Brazil
}

\date{Accepted XXX. Received YYY; in original form ZZZ}

\pubyear{2015}

\begin{document}
\label{firstpage}
\pagerange{\pageref{firstpage}--\pageref{lastpage}}
\maketitle

\begin{abstract}
Fanaroff--Riley type 0 (FR0) radio galaxies host anomalously 
compact jets whose disruption mechanism and high-energy emission 
remain poorly understood. We combine three-dimensional 
relativistic hydrodynamical (RHD) simulations with broadband 
spectral energy distribution (SED) modeling from radio to TeV 
energies, focusing on LEDA~55267 and LEDA~58287. Our simulations 
show that recollimation shocks trigger hydrodynamical 
instabilities that drive turbulence and rapid deceleration, 
preventing the jets from propagating beyond a few tens of 
parsecs and reproducing the observed compact radio morphology. 
Leptonic SED models adequately describe the observed emission 
up to GeV energies, but when simulated CTAO observations are included, statistical model comparison indicates strong evidence in favor of lepto-hadronic scenarios at TeV energies for both sources, a result that should be interpreted as a theoretical prediction to be tested by future observations. A leptonic analysis of the simulations reveals magnetized emitting regions 
with plasma beta parameters $\beta_{\rm p} \sim 
10^{-5}$--$10^{-3}$, orders of magnitude below values reported 
for extended FRI jets, consistent with jets retaining the 
magnetization inherited from the launching region and providing 
a natural physical link between the compact jet dynamics and 
the lepto-hadronic emission.
\end{abstract}



\section{Introduction}

Supermassive black holes accreting matter give rise to what are known as active galactic nuclei (AGNs), which are capable of producing relativistic jets that can be collimated and extend up to megaparsec scales~\citep{Blandford_2019}. According to the unified model 
of AGNs~\citep{Urry_1995}, radio galaxies (RGs) are sources whose jets, characterized by strong radio emission, are not aligned with the line of sight, making them the dominant class of jetted AGNs. These sources are traditionally classified as Fanaroff--Riley type I (FRI) or Fanaroff--Riley type II (FRII) according to their radio morphology~\citep{Fanaroff_1974}. FRI sources exhibit more diffuse and less powerful jets with edge-darkened structures, whereas FRII sources are more powerful and display edge-brightened morphologies with prominent hotspots. Within these jets, particles are accelerated to relativistic velocities, enabling electromagnetic interactions that can 
produce multimessenger particles, including neutrinos and ultra-high-energy cosmic rays (UHECRs)~\citep[see, e.g.,][]{Dermer_2009, Aartsen_2018, Tavecchio_2018, Abassi_2022, 
Buson_2022}, as well as high-energy gamma-ray emission~\citep[see, e.g.,][]{Abdo_2009, Abdo_2010, Inoue_2011, Dimauro_2014, Angioni_2019, Stecker_2019, Harvey_2020, 
Rulten_2020}.

Although FRI and FRII radio galaxies have been extensively studied, a third class of radio galaxies, referred to as Fanaroff--Riley type~0 (FR0), was introduced to describe sources hosting anomalously compact jets that lack the extended radio structures characteristic of their classical counterparts~\citep{Ghisellini_2011, Baldi_2015, Baldi_2018, Baldi_2019b, Capetti_2020b}. Recent reviews have consolidated the current understanding of FR0 radio galaxies, highlighting their compact morphology, high core dominance, and lack of extended radio emission, despite sharing similar host and nuclear
properties with FRIs \citep{Baldi_2023}. These properties indicate that FR0 jets undergo early disruption or inefficient propagation, making them key targets for studying the connection between jet dynamics and high-energy emission. According to the FR0CAT catalog~\citep{Baldi_2018}, FR0s host black holes with masses in the range $10^{7.4}$--$10^{9.0}\,\mathrm{M_{\sun}}$ and are predominantly early-type, low-excitation galaxies found at redshifts $z < 0.05$. Remarkably, they are about five times more numerous than FRIs and approximately twenty times more numerous than FRIIs in the 
local Universe, making them the dominant population of jetted AGNs at low redshift. Given this abundance, FR0s may contribute significantly to the diffuse backgrounds of cosmic rays and neutrinos~\citep{Merten_2021, Lundquist_2025}, making their characterization a problem of broad relevance to high-energy astrophysics. FR0s have 
1.4 GHz radio luminosities in the range $10^{38.19}$-$10^{40.27}\,\mathrm{erg\,s^{-1}}$ and share several properties with FRIs, such as their X-ray luminosity in the 2--10 keV 
band~\citep{Torresi_2018}. Despite these similarities, FR0 jets are confined to scales of less than 5 kpc, and no consensus explanation for this striking compactness has yet emerged. Among the proposed mechanisms, low black-hole spin, prograde accretion, and 
advection-dominated flows have been explored as possible causes for the suppression of jet growth~\citep{Garofalo_2010, Garofalo_2019, Garofalo_2024, Torresi_2022}. A summary 
of the main observational properties of the three classes is presented in Table~\ref{tab:properties_fr0}, based on the FR0CAT, FRICAT, and FRIICAT catalogs~\citep{Baldi_2018, Capetti_2017a, Capetti_2017b}.

\begin{table}
    \centering
    \setlength\tabcolsep{0.16cm}
    \caption{Summary of the main properties of FR0, FRI, and FRII radio galaxies as 
    reported in the FR0CAT, FRICAT, and FRIICAT catalogs 
    \citep{Baldi_2018, Capetti_2017a, Capetti_2017b}.}
    \label{tab:properties_fr0}
    \begin{tabular}{lccc}
        \hline\hline
        Property & FR0CAT & FRICAT & FRIICAT \\
        \hline
        Number & 108 & 219 & 122 \\
        Redshift & $z < 0.05$ & $z < 0.15$ & $z < 0.15$ \\
        $L_{\rm r}$ [erg s$^{-1}$] 
            & $10^{38.19} - 10^{40.27}$ 
            & $10^{39.50} - 10^{41.30}$ 
            & $10^{39.50} - 10^{42.50}$ \\
        $M_{\rm r}$ 
            & $[-23, -21]$ 
            & $[-24, -21]$ 
            & $[-24, -20]$ \\
        Jet size [kpc] & $r < 5$ & $r > 30$ & $r > 30$ \\
        $M_{\rm BH}$ [$\rm M_{\sun}$] 
            & $10^{7.4} - 10^{9.0}$ 
            & $10^{8.0} - 10^{9.5}$ 
            & $10^{6.5} - 10^{9.5}$ \\
        Host galaxy   & Early-type & Early-type & Early-type \\
        Excitation    & Low        & Low        & High/Low   \\
        \hline\hline
    \end{tabular}
\end{table}

Other approaches to studying FR0 radio galaxies are based on the morphological analysis of their radio jets. These studies show that FR0s are extremely compact sources whose jets have moderately relativistic velocities and may undergo changes in direction, while 
also suggesting that these sources host low-luminosity AGNs capable of driving galactic winds~\citep[see, e.g.,][]{Baldi_2019b, Baldi_2025, Roy_2021, Giovannini_2023, Chilufya_2024}. A natural mechanism for suppressing jet propagation on kiloparsec scales is the development of recollimation shocks at the jet boundary, which can trigger 
hydrodynamical instabilities that promote turbulence, mixing with the ambient medium, and rapid deceleration of the flow. Complementary numerical investigations have explored this scenario in detail.~\citet{Costa_2024} and~\citet{Costa_2026} employ relativistic 
hydrodynamical (RHD) simulations to demonstrate how recollimation shocks can inhibit jet evolution on large scales. Building on these works,~\citet{Boula_2025} use simulations in the relativistic magnetohydrodynamics (RMHD) regime to investigate the role of the magnetic field in jet stability, also accounting for the presence of 
recollimation shocks.~\citet{Lalakos_2024} study jet formation by considering the black hole spin and, using general relativistic magnetohydrodynamics (GRMHD), propose that the infall of gas with negligible angular momentum onto rapidly rotating black holes 
prevents the formation of a stable accretion disc, rendering the jets intrinsically unstable. Meanwhile,~\citet{Borodina_2025} employ hydrodynamical (HD) simulations of low- and intermediate-power jets propagating through a turbulent interstellar medium (ISM), showing that such turbulence can suppress jet growth and evolution. Taken 
together, these recent studies indicate that FR0s play a relevant role in the physics of relativistic jets and may constitute promising environments for studying particle acceleration and the origin of UHECRs.

Although the jet power of FR0 radio galaxies is lower than that of FRI and FRII sources, this class of radio galaxies has been proposed as a promising source of neutrino emission~\citep{Tavecchio_2018} and as a potential acceleration site of UHECRs~\citep{Merten_2021, Lundquist_2025}. Indeed, very-high-energy gamma-ray emission has 
already been associated with FR0 jets.~\citet{Grandi_2016} associated gamma-ray sources from the 3FGL \textit{Fermi}-LAT catalog~\citep{Acero_2015} with the FR0 radio galaxy Tol~1326-379; however, no gamma-ray emission from this source was reported in the 4FGL \textit{Fermi}-LAT catalog, and it therefore remains unclear whether Tol~1326-379 is indeed a gamma-ray emitter~\citep{Fu_2022}.~\citet{Pannikkote_2023} cross-matched the 4FGL \textit{Fermi}-LAT catalog with radio and optical surveys and identified seven radio galaxies as candidate FR0 sources. More conclusively,~\citet{Paliya_2021} reported GeV gamma-ray emission detected in the direction of two FR0 radio galaxies from the FR0CAT catalog, hereafter referred to as LEDA~55267 and LEDA~58287. These two sources are particularly well-suited for detailed investigation, as they are both located at low redshift ($z < 0.05$), exhibit parsec-scale jet-like radio structures confirmed by high-resolution VLBI observations~\citep{Giovannini_2023}, and display moderately relativistic jet velocities~\citep{Baldi_2019b} that make them amenable to RHD modeling. This detection opens a broad window for investigation, since FR0s are the most numerous jetted AGN population in the local Universe and may therefore represent a significant class of extragalactic gamma-ray sources detectable on Earth. More broadly, gamma-ray emission from low-power compact AGN has
recently attracted increasing attention, with several works exploring
particle acceleration and radiative processes in such systems~\citep[e.g.,][]{Cao_2024, Bronzini_2024a, Bronzini_2024b, Jiang_2026}. These studies indicate that even low-power jets may contribute significantly to the diffuse gamma-ray background, reinforcing the relevance of FR0 sources as potential high-energy emitters. Furthermore,~\citet[][hereafter BR23]{Boughelilba_2023} and~\citet{Khatiya_2024} explored the gamma-ray emission from LEDA~55267 and LEDA~58287 through spectral energy distribution (SED) modeling, considering lepto-hadronic and leptonic scenarios, respectively, reinforcing the idea that these sources are capable of accelerating particles to very high energies. Preliminary results combining RHD simulations with CTAO 
prospects~\citep[CTAO;][]{CTAO_consortium} for these two sources were recently presented in~\citet{Cardoso_2026_HEPRO}, and the present work provides a full account of those investigations.

The combination of compact, low-power jets and possible high-energy gamma-ray emission makes FR0 radio galaxies particularly intriguing, as it challenges the classical particle-acceleration scenarios associated with more powerful radio galaxies. A key difficulty in distinguishing between leptonic and lepto-hadronic emission models is that both scenarios produce degenerate predictions at GeV energies, while diverging significantly at TeV energies, where hadronic contributions become dominant. This makes next-generation Cherenkov observatories essential tools for breaking this degeneracy. In this work, we investigate LEDA~55267 and LEDA~58287 by combining two complementary approaches. On one hand, we exploit the most recent gamma-ray data from the 4FGL \textit{Fermi}-LAT catalog~\citep{Abdollahi_2020, Ballet_2023} together with simulated observations from the Cherenkov Telescope Array Observatory~\citep[CTAO;][]{CTAO_consortium} to model the broadband SEDs and evaluate the detectability of these sources at TeV energies. On the other hand, we present three-dimensional RHD simulations performed with the PLUTO code~\citep{Mignone_2007} to reproduce the observed compact morphology of these sources. Crucially, the synthetic emission maps derived from the simulations serve as an independent consistency check for the emitting regions inferred from the SED analysis, linking jet dynamics directly to the observed radiative properties. We consider both leptonic and lepto-hadronic scenarios to constrain the physical parameters of both sources.

This paper is organized as follows. Section~\ref{sec:sources} 
presents the observational properties of LEDA~55267 and 
LEDA~58287 that motivate and constrain our modeling. 
Section~\ref{sec:methods} describes the methods adopted, covering 
the jet simulation setup, the SED modeling framework, and the 
CTAO observational configuration. Section~\ref{sec:results} 
presents the results, including the jet dynamics and disruption, 
the synthetic emission maps, the broadband SED modeling under 
leptonic and lepto-hadronic scenarios, and the plasma 
magnetization analysis. Section~\ref{sec:conclusions} discusses
the physical interpretation of the results and their broader 
implications for FR0 jets as sites of high-energy particle 
acceleration and summarizes the main 
conclusions of this work.

\section{Sources and Observational Constraints}
\label{sec:sources}

\begin{figure}
    \centering
    \includegraphics[width=0.95\columnwidth]{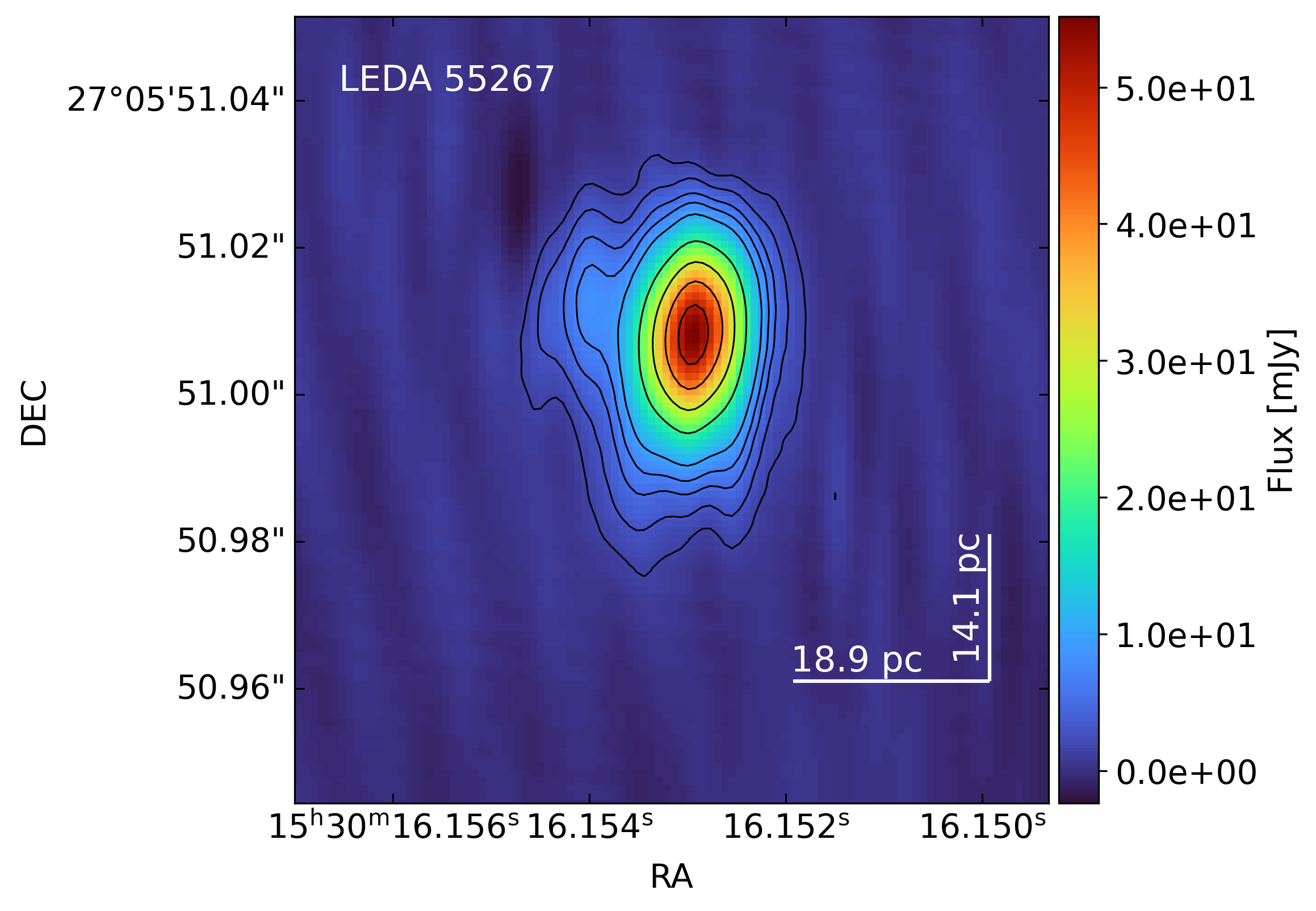} \\
    \includegraphics[width=0.95\columnwidth]{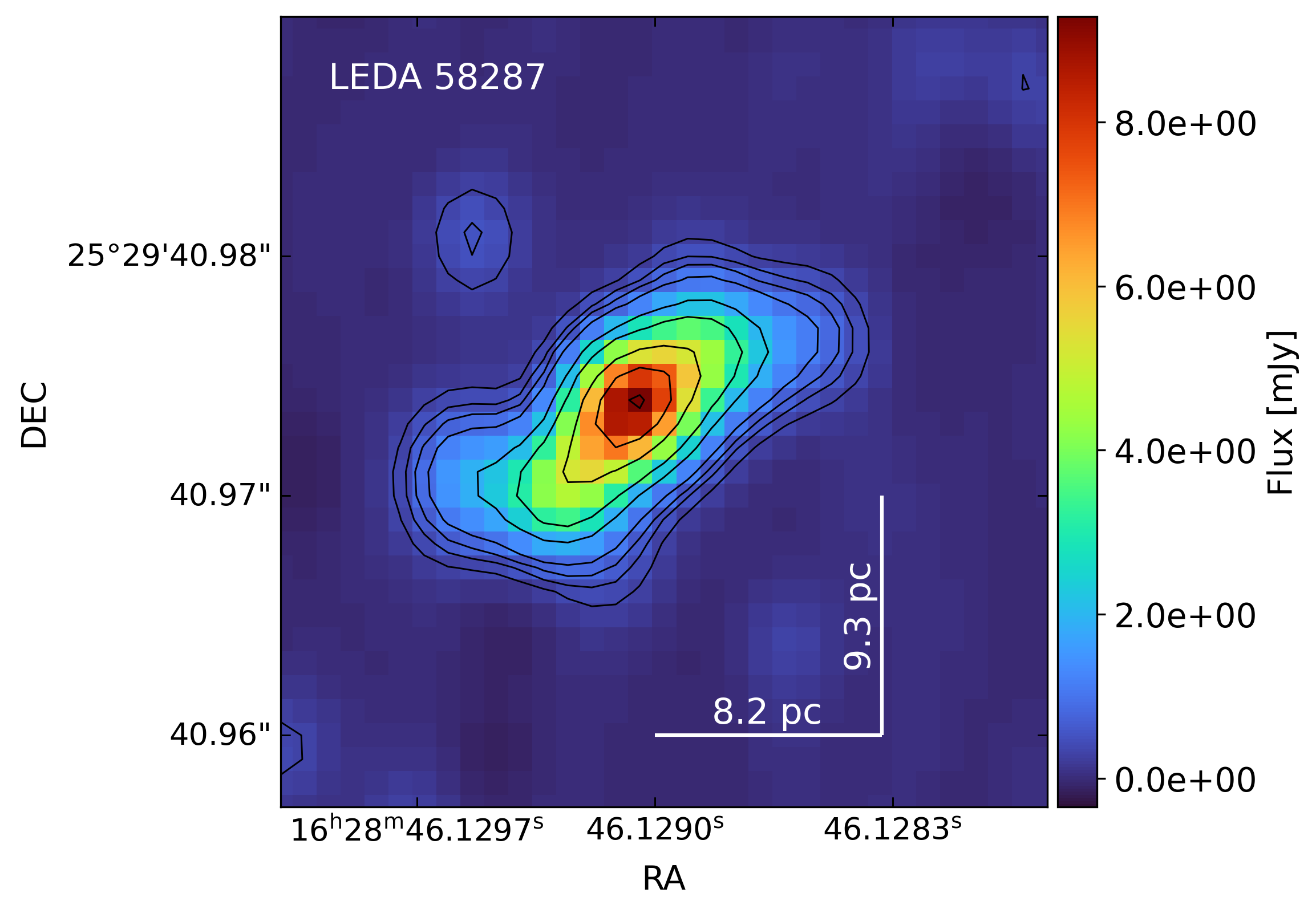}
    \caption{Radio maps of LEDA~55267 and LEDA~58287. The upper panel shows LEDA~55267 observed with the EVN at 1.7~GHz, with a peak flux density of $5.5 \times 10^{-2}$~Jy\,beam$^{-1}$ and contour levels of $(1.5,\, 3.0,\, 5.0,\, 7.0,\, 10.0,\, 20.0,\, 30.0,\, 40.0,\, 50.0) \times 10^{-3}$~Jy\,beam$^{-1}$. The lower panel shows LEDA~58287 observed with the VLBA at 1.5~GHz, with a peak flux density of $9.3 \times 10^{-3}$~Jy\,beam$^{-1}$ and contour levels of $(-0.3,\, 0.3,\, 0.5,\, 0.7,\, 1.0,\, 2.0,\, 3.0,\, 5.0,\, 7.0,\, 9.0) \times 10^{-3}$~Jy\,beam$^{-1}$. Both sources show compact morphologies consistent with jet structures on parsec scales. Radio maps adapted from~\citet{Giovannini_2023}.}
    \label{fig:radio_ledas}
\end{figure}

The radio galaxies LEDA~55267 and LEDA~58287 have been observed at multiple radio frequencies using the VLA, VLBA, VLBI, and EVN \citep{Baldi_2019b, Giovannini_2023}. Both sources appear unresolved even at high angular resolution, confirming their extremely compact nature. LEDA~55267 exhibits a significant increase in flux density 
across epochs, suggesting variability, and on parsec scales presents a dominant core with a morphology consistent with a jet structure \citep{Giovannini_2023}. LEDA~58287 similarly shows a slight increase in flux density and, on parsec scales, displays outer 
regions rotated with respect to the inner region, which is interpreted as evidence for the presence of a jet \citep{Giovannini_2023}. The radio maps of both sources, observed with the EVN at 1.7~GHz and with the VLBA at 1.5~GHz for LEDA~55267 and LEDA~58287, respectively, are presented in Figure~\ref{fig:radio_ledas}.

As shown in Figure~\ref{fig:radio_ledas}, both sources display jet-like structures on scales of the order of tens of parsecs and have been detected in gamma-rays, making them well-suited targets for combined dynamical and radiative modeling. These observational constraints directly inform the setup of our jet simulations, which are designed to reproduce both the physical dimensions and the compact morphology observed at radio wavelengths.

\section{Methods}
\label{sec:methods}

To investigate the physical properties of LEDA~55267 and 
LEDA~58287, we adopt three complementary methodological 
approaches that are closely coupled. First, we 
perform three-dimensional RHD simulations to reproduce the 
observed compact jet morphology and obtain the dynamical 
properties of the flow, including the thermal pressure 
distribution that will later be used to estimate the magnetic 
field strength. Second, we model the broadband SEDs of both 
sources from radio to gamma-ray energies, considering leptonic 
and lepto-hadronic particle populations, using the synthetic 
emission maps derived from the simulations for a consistency check on the physical extent of the emitting regions. Third, we simulate CTAO observations to assess the detectability of 
both sources at TeV energies, and to evaluate the diagnostic 
power of future observations in discriminating between emission 
scenarios. Each of these approaches is described in the following subsections.

\subsection{Jet simulation setup}
\label{sec:jet_parameters}

For the simulation of relativistic jets, we used the public PLUTO code 
\citep{Mignone_2007} in the RHD regime, performing a three-dimensional 
simulation in Cartesian coordinates. The HLLC Riemann solver and second-order Runge--Kutta (RK2) time stepping were employed in a computational domain with a grid of $350 \times 350 \times 1000$ cells, covering the physical domain $[-17.5, 17.5] \times [-17.5, 17.5] \times [2.5, 150]$~pc. Outflow boundary conditions were applied on all domain faces, with the exception of the lower $z$-boundary, where the jet injection conditions are imposed. To track the jet material throughout the simulation, a passive scalar tracer field $\phi$ is advected alongside the fluid, initialized to $\phi = 1$ inside the jet cone and $\phi = 0$ in the external medium. This tracer allows us to distinguish jet material from the ambient medium at all times and to 
identify regions of mixing. The equations governing the RHD regime express the conservation of mass, momentum, and energy~\citep{Landau_1959}, given respectively by:
\begin{equation}
    \frac{\partial D}{\partial t} + \nabla \cdot (D\,\mathbf{v}) = 0,
\end{equation}
\begin{equation}
    \frac{\partial \mathbf{M}}{\partial t} + 
    \nabla \cdot (\mathbf{M} \otimes \mathbf{v} + p\,\mathbf{I}) = 0,
\end{equation}
\begin{equation}
    \frac{\partial E}{\partial t} + \nabla \cdot \mathbf{M} = 0,
\end{equation}
where $D = \rho\Gamma$ is the rest-frame mass density, 
$\mathbf{M} = \rho h \Gamma^{2}\mathbf{v}$ is the momentum density, and $E = \rho h \Gamma^{2} - p$ is the total energy density. Here, $\rho$ is the rest-mass density, $\Gamma$ is the Lorentz factor, $\mathbf{v}$ is the fluid velocity, $p$ is the thermal pressure, and $h = 1 + e + p/\rho$ is the specific enthalpy, where $e$ is the specific internal energy density. For the equation of state (EOS), we adopt the Taub--Matthews (TM) EOS~\citep{Mignone_2005}, for which the specific enthalpy is given by:
\begin{equation}
    h = \frac{5}{2}\Theta + \sqrt{\frac{9}{4}\Theta^{2} + 1},
    \label{eq:enthalpy}
\end{equation}
\noindent where $\Theta = p/(\rho c^{2})$, and we assume $c = 1$.
Since FR0 jets are observed to be suppressed on kiloparsec scales, we base our simulation on the recollimation shock instability model proposed by~\citet{Costa_2024}. In this framework, recollimation shocks trigger hydrodynamical instabilities that prevent the jet from further evolving, reproducing the compact morphology observed in FR0 sources. The jet is launched with an opening angle $\theta_{\rm j} = 0.2$~rad, and at $t = 0$, within the conical region $(\sqrt{x^2 + y^2})/z < 0.2$, the jet velocity, density, and pressure are initialized according to
\begin{equation}
    \mathbf{v} = \sqrt{1 - \frac{1}{\Gamma_{\rm j}^{2}}}\,\frac{\mathbf{r}}{R},
\end{equation}
\begin{equation}
    \rho_{\rm j}(x, y, z, 0) = \rho_{\rm j}(0, 0, z_{0}, 0) 
    \left(\frac{R}{R_{0}}\right)^{-2},
\end{equation}
\begin{equation}
    p_{\rm j}(x, y, z, 0) = p_{\rm j}(0, 0, z_{0}, 0) 
    \left(\frac{R}{R_{0}}\right)^{-2\gamma},
    \label{eq:pressure_jet_init}
\end{equation}
where $z_{0}$ is the jet launching distance, $\gamma$ is the 
adiabatic index, and $R = \sqrt{x^{2} + y^{2} + z^{2}}$.
In the region external to the jet, where $(\sqrt{x^2 + y^2})/z > 0.2$, the density and pressure of the ambient medium follow power-law profiles in $z$,
\begin{equation}
    \rho_{\rm ext}(z, 0) = \rho_{\rm ext}(z_{0}, 0) 
    \left(\frac{z}{z_{0}}\right)^{-\eta},
\end{equation}
\begin{equation}
    p_{\rm ext}(z, 0) = p_{\rm ext}(z_{0}, 0) 
    \left(\frac{z}{z_{0}}\right)^{-\eta},
\end{equation}
where we adopt $\eta = 0.5$. As shown by~\citet{Costa_2024}, the precise value of $\eta$ does not critically affect the jet evolution, provided that $\eta < 2$. Since FR0 radio galaxies share several properties with FRIs, we adopt external medium pressure and density values comparable to those inferred for the FRI radio galaxy M87, namely $p_{\rm ext} = 1.5 \times 10^{-9}$~dyne~cm$^{-2}$ and $\rho_{\rm ext} = 1.0$~cm$^{-3}$~\citep{Sparks_1996, Stawarz_2006, Dainotti_2012, 
Boizelle_2025}. We note that this choice represents an 
approximation, since the interstellar medium conditions in the 
host galaxies of LEDA~55267 and LEDA~58287 are not individually 
constrained by X-ray observations. However, these values are 
consistent with the range of thermal pressures reported for 
early-type galaxies at low redshift~\citep[see, e.g.,][]{Canizares_1987, Boroson_2011, Torresi_2018}, and \citet{Costa_2024} show that the jet disruption behavior is not sensitive to moderate variations in the external medium parameters. Following \citet{Costa_2024}, we adopt a light jet with density $\rho_{\rm j} = 10^{-4}$~cm$^{-3}$, well below that of the surrounding medium. The initial jet pressure is 
derived from the radio luminosity through the relation
\begin{equation}
    L_{\rm j} = \pi(z_{0}\,\theta)^{2}\,v_{{\rm j},z}\,
    \rho_{{\rm j},0}\,c^{2}\,h_{{\rm j},0}\,\Gamma^{2}.
    \label{eq:jet_luminosity}
\end{equation}

\begin{table}
    \centering
    \caption{RHD simulation setup parameters for the jet and external medium.}
    \label{tab:setup_simulation}
    \begin{tabular}{llcc}
        \hline\hline
        Component & Quantity & Value & Units \\
        \hline
        \multirow{6}{*}{Jet}
            & Density         & $1.7 \times 10^{-28}$ & g\,cm$^{-3}$ \\
            & Velocity        & $0.7\,c$              & ---          \\
            & Lorentz factor  & $1.4$                 & ---          \\
            & Pressure        & $8.1 \times 10^{-8}$  & dyne\,cm$^{-2}$ \\
            & Opening angle   & $0.2$                 & rad          \\
            & $L_{\rm r}$     & $4.9 \times 10^{38}$  & erg\,s$^{-1}$ \\
        \hline
        \multirow{2}{*}{Environment}
            & Density         & $1.7 \times 10^{-24}$ & g\,cm$^{-3}$ \\
            & Pressure        & $1.5 \times 10^{-9}$  & dyne\,cm$^{-2}$ \\
        \hline\hline
    \end{tabular}
\end{table}

According to the FR0CAT catalog~\citep{Baldi_2018}, the radio 
luminosities of LEDA~55267 and LEDA~58287 are $L_{\rm r} = 
10^{38.69}$~erg~s$^{-1}$ and $L_{\rm r} = 10^{39.15}$~erg~s$^{-1}$, respectively. Since the two values differ by less than a factor of three and are of the same order of magnitude, we adopt the luminosity of LEDA~55267 as the reference to estimate the initial jet pressure, obtaining $p_{\rm j} = 8.1 \times 10^{-8}$~dyne~cm$^{-2}$ from Equations~\ref{eq:enthalpy} and~\ref{eq:jet_luminosity}. We verified that varying the jet pressure within this factor of three does not qualitatively alter the disruption dynamics described in Section~\ref{sec:simulation_results}. We adopt an initial jet velocity of $v_{\rm j} = 0.7\,c$, representative of the conditions close to the jet launching region. While VLBI observations and theoretical studies of FR0 radio galaxies indicate mildly relativistic bulk speeds on parsec scales~\citep[$\lesssim 0.5\,c$; e.g.][]{Baldi_2015, Baldi_2019b, Cheng_2018, Capetti_2020a, Cheng_2021, Giovannini_2023}, these measurements likely probe already decelerated flows rather than the pristine launch region. In our simulations, the jet undergoes rapid deceleration through recollimation shocks and hydrodynamical instabilities (Section~\ref{sec:simulation_results}), so that the bulk velocities recovered over most of the jet extent are fully consistent with the observational constraints. Adopting an initial velocity below $\sim 0.7\,c$ would correspond to a less powerful jet with lower momentum flux, which would be even more easily disrupted by the same instabilities; in order for such a slower flow to evolve self-consistently with the same external medium, the ambient density would have to be reduced below the values inferred for early-type galaxy hosts, compromising the
physical consistency of the setup. We note that the initial parameters of our simulation differ from those of \citet{Costa_2024} in three main
respects: we adopt $\Gamma = 1.4$ rather than $\Gamma = 10$, in
agreement with the moderately relativistic speeds inferred for FR0 sources; the initial jet pressure is derived directly from the observed
1.4~GHz radio luminosity through Equation~\ref{eq:jet_luminosity}, rather than being imposed as a free numerical parameter; and the external medium properties are anchored to observationally motivated values for FRI hosts such as M87. The complete set of simulation parameters is summarized in Table~\ref{tab:setup_simulation}.

\subsection{SED modeling framework}
\label{sec:SED}

To construct the broadband SEDs of LEDA~55267 and LEDA~58287, we 
compile multiwavelength data from radio to gamma-ray energies 
available in~\citetalias{Boughelilba_2023}, excluding the host 
galaxy contribution in order to isolate the non-thermal synchrotron and inverse Compton emission. These data are supplemented with the most recent flux measurements from the 4FGL \textit{Fermi}-LAT catalog~\citep{Abdollahi_2020, Ballet_2023}.

The particle populations responsible for the radio and gamma-ray 
emission are assumed to be co-spatial, consistent with the modeling framework of~\citet{Khatiya_2024}. External radiation fields capable of driving significant particle--photon interactions are not expected for FR0 sources, since the internal synchrotron radiation field of the jet greatly exceeds any ambient external field. Nevertheless, we include the cosmic microwave background (CMB) as an additional seed photon field for inverse Compton scattering. The synchrotron self-Compton (SSC) emission is assumed to originate in an effective emitting region located between $10^{3}$ and $10^{5}$ gravitational radii $r_{\rm g}$ from the central black hole, following the scenario proposed 
by~\citet{Khatiya_2024}. This range corresponds to physical 
scales of the order of sub-parsec to tens of parsecs for the 
black hole masses typical of FR0 sources~\citep{Baldi_2018}, consistent with the parsec-scale jet structures resolved by~\citet{Giovannini_2023} for both LEDA~55267 and LEDA~58287. This emitting region, therefore, represents an average parametrization of the physical conditions responsible for the observed non-thermal emission, grounded in the observational constraints available for both sources. Within this region, the spectral photon density of the synchrotron radiation field is given by
\begin{equation}
    n_{\rm syn}(E) = \frac{2.24}{4\pi R_{\rm jet}^{2}\,c}\,L_{\rm syn}(E),
    \label{eq:photon_density_syn}
\end{equation}
\noindent where $E$ is the photon energy, $R_{\rm jet}$ is the radius of the emitting region, and $L_{\rm syn}(E)$ is the synchrotron luminosity. The numerical factor 2.24 accounts for geometric effects under the assumption of a spherically symmetric emission region~\citep{Atoyan_1996}.

The SEDs are fitted using the publicly available code 
\texttt{NAIMA}~\citep{naima_Zabalza_2015}, which implements 
analytical models for synchrotron and inverse Compton emission 
following~\citet{Synchrotron_Aharonian_2010} and~\citet{Compton_Khangulyan_2014}, respectively. The energy 
distributions of relativistic electrons and protons are modeled as 
an exponential cutoff power-law (ECPL) of the form
\begin{equation}
    f_{\rm e,p}(E) = A_{\rm e,p} \left(\frac{E}{E_{0}}\right)^{-\alpha_{\rm e,p}} 
    \exp\left(-\frac{E}{E_{\rm cut,e,p}}\right),
    \label{eq:ECPL_sed}
\end{equation}
\noindent where $A_{\rm e,p}$ is the normalization at the reference energy $E_{0}$, $\alpha_{\rm e,p}$ is the spectral index, and $E_{\rm cut,e,p}$ is the cutoff energy for electrons and protons, respectively.

\subsection{CTAO observational}
\label{sec:ctao}

To assess the detectability of LEDA~55267 and LEDA~58287 at TeV 
energies, we simulate CTAO observations using the one-dimensional 
(1D) ON/OFF observational technique implemented in the 
\texttt{Gammapy} software~\citep{gammapy_2023, 
acero_2025_14760974}\footnote{\href{https://gammapy.org}{https://gammapy.org}}. The ON region is defined as a circular aperture of radius $0.5^{\circ}$ centered on each source, covering the GeV--TeV energy range of interest. Flux measurements from the 4FGL \textit{Fermi}-LAT catalog~\citep{Abdollahi_2020, Ballet_2023} are included to anchor the spectral models at GeV energies and ensure continuity across the full energy range covered by the simulation.

The annual visibility of both sources from the CTAO North and South sites was computed for 2026 at zenith angles of $20^{\circ}$, $40^{\circ}$, and $60^{\circ}$, and is listed in 
Table~\ref{tab:visibility_CTAO}. Neither source is accessible from 
the CTAO South at zenith angles below $60^{\circ}$, whereas both 
sources accumulate between 556 and 580~hours of visibility per year from the CTAO North at all three zenith angles. We therefore adopt the CTAO North at a zenith angle of $20^{\circ}$ as the reference observational configuration throughout this work.

\begin{table}
    \centering
    \caption{Annual visibility of LEDA~55267 and LEDA~58287 from 
    the CTAO South and North arrays at zenith angles of $20^{\circ}$, 
    $40^{\circ}$, and $60^{\circ}$. Dashes indicate configurations 
    at which the source is not accessible.}
    \label{tab:visibility_CTAO}
    \begin{tabular}{llcc}
        \hline\hline
        Source & CTAO array & Zenith angle & Visibility [h] \\
        \hline
        \multirow{6}{*}{LEDA~55267}
            & South & $20^{\circ}$ & --- \\
            & South & $40^{\circ}$ & --- \\
            & South & $60^{\circ}$ & 1190.50 \\
            & North & $20^{\circ}$ & 556.50 \\
            & North & $40^{\circ}$ & 561.50 \\
            & North & $60^{\circ}$ & 580.00 \\
        \hline
        \multirow{6}{*}{LEDA~58287}
            & South & $20^{\circ}$ & --- \\
            & South & $40^{\circ}$ & --- \\
            & South & $60^{\circ}$ & 1230.50 \\
            & North & $20^{\circ}$ & 560.00 \\
            & North & $40^{\circ}$ & 559.50 \\
            & North & $60^{\circ}$ & 574.00 \\
        \hline\hline
    \end{tabular}
\end{table}

The Instrument Response Functions \citep[IRFs;][]{CTAO_2021_IRF} 
provided by the CTAO Consortium, optimized for an observation time 
of 50~hours, are used to generate the sensitivity curve. Detection 
is required to satisfy a minimum significance of $5\sigma$ with at 
least 10 detected photons, and a systematic background uncertainty 
of 0.05 is assumed throughout. An observation time of 200~hours is 
adopted for each source, from which we derive the CTAO sensitivity 
curve and the simulated gamma-ray flux points.

To account for gamma-ray attenuation by the Extragalactic Background Light (EBL), we apply the model of~\citet{Saldana-Lopez_2021}. The EBL attenuates high-energy gamma-rays through pair production with intergalactic photons, with the effect increasing with source distance. This attenuation is incorporated into all spectral models to avoid overestimating the intrinsic source emission.

\section{Results}
\label{sec:results}

This section presents the results of our combined dynamical and 
radiative investigation of LEDA~55267 and LEDA~58287, organized 
to reflect the physical chain of reasoning that connects the 
three methodological approaches described in 
Section~\ref{sec:methods}. We begin by presenting the jet 
dynamics and disruption mechanism emerging from the RHD 
simulations, followed by the synthetic synchrotron and inverse 
Compton emission maps that link the simulation results to the 
observed radio morphology. We then present the broadband SED 
modeling, incorporating the most recent \textit{Fermi}-LAT 
data~\citep{Abdollahi_2020, Ballet_2023} together with simulated 
CTAO observations at TeV energies, and assess the statistical 
preference between leptonic and lepto-hadronic scenarios through 
the Bayesian Information Criterion. Finally, we use the thermal 
pressures extracted from the simulations to estimate the plasma 
magnetization of the emitting regions through the $\beta_{\rm p}$ 
parameter, providing an independent consistency check between the 
hydrodynamical and radiative analyses. Taken together, these 
results build a coherent physical picture in which the compact 
morphology, the high-energy emission, and the magnetization state 
of FR0 jets can be understood as manifestations of the same 
underlying jet dynamics.

\subsection{Jet dynamics and disruption}
\label{sec:simulation_results}

\begin{figure*}
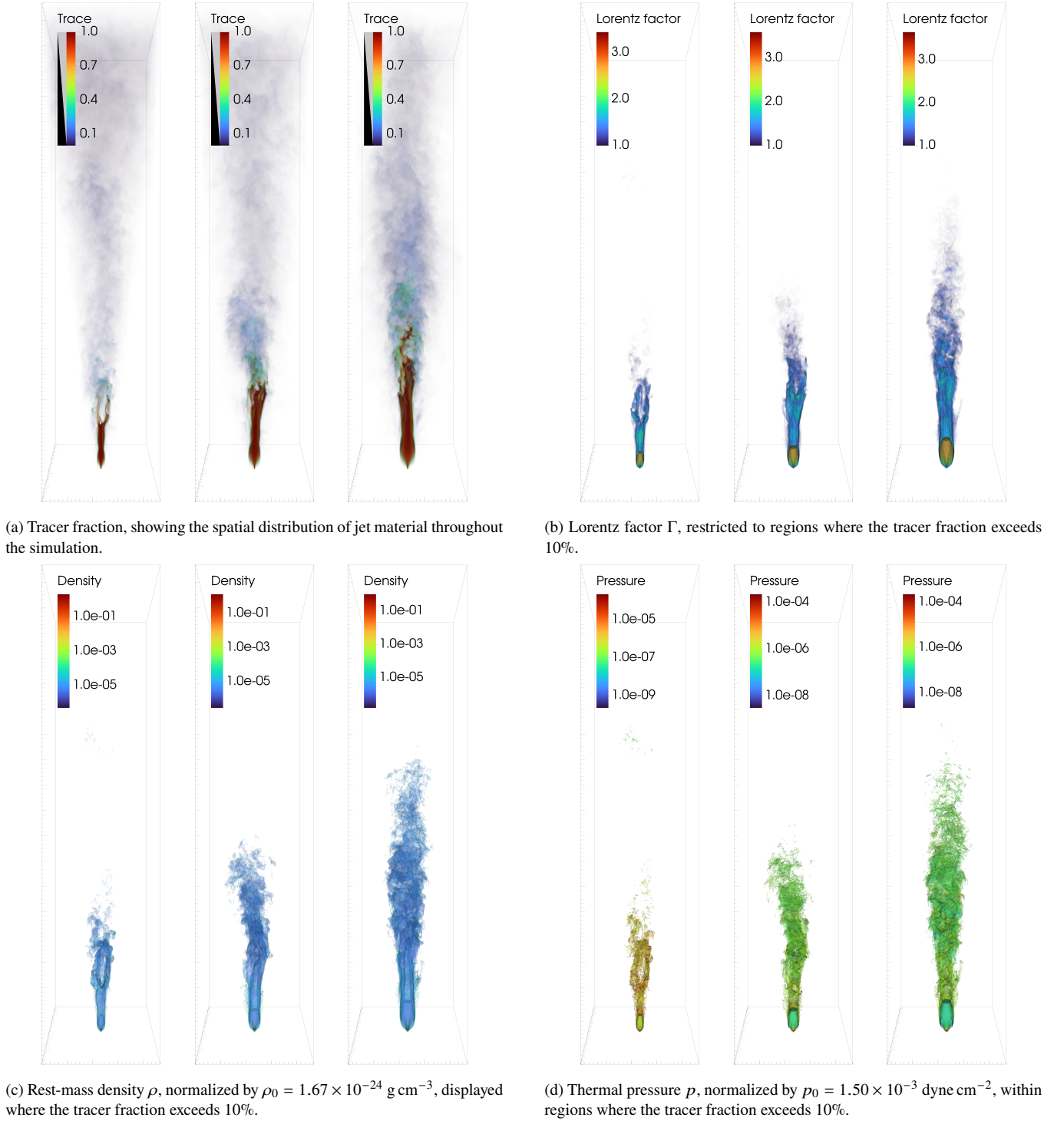

    \centering
    
    \begin{subfigure}{0.48\textwidth}
        \centering
        \includegraphics[width=0.3\linewidth]{figures_arxiv/trace0000.jpg}
        \includegraphics[width=0.3\linewidth]{figures_arxiv/trace0001.jpg}
        \includegraphics[width=0.3\linewidth]{figures_arxiv/trace0002.jpg}
        \caption{Tracer fraction, showing the spatial distribution 
        of jet material throughout the simulation.}
        \label{fig:simulations_a}
    \end{subfigure}
    \hfill
    \begin{subfigure}{0.48\textwidth}
        \centering
        \includegraphics[width=0.3\linewidth]{figures_arxiv/gamma0000.jpg}
        \includegraphics[width=0.3\linewidth]{figures_arxiv/gamma0001.jpg}
        \includegraphics[width=0.3\linewidth]{figures_arxiv/gamma0002.jpg}
        \caption{Lorentz factor $\Gamma$, restricted to regions where the tracer fraction exceeds 10\%.}
        \label{fig:simulations_b}
    \end{subfigure}
    \vfill
    \begin{subfigure}{0.48\textwidth}
        \centering
        \includegraphics[width=0.3\linewidth]{figures_arxiv/rho0000.jpg}
        \includegraphics[width=0.3\linewidth]{figures_arxiv/rho0001.jpg}
        \includegraphics[width=0.3\linewidth]{figures_arxiv/rho0002.jpg}
        \caption{Rest-mass density $\rho$, normalized by 
        $\rho_{0} = 1.67 \times 10^{-24}$~g\,cm$^{-3}$, displayed where 
        the tracer fraction exceeds 10\%.}
        \label{fig:simulations_c}
    \end{subfigure}
    \hfill
    \begin{subfigure}{0.48\textwidth}
        \centering
        \includegraphics[width=0.3\linewidth]{figures_arxiv/prs0000.jpg}
        \includegraphics[width=0.3\linewidth]{figures_arxiv/prs0001.jpg}
        \includegraphics[width=0.3\linewidth]{figures_arxiv/prs0002.jpg}
        \caption{Thermal pressure $p$, normalized by 
        $p_{0} = 1.50 \times 10^{-3}$~dyne\,cm$^{-2}$, within regions 
        where the tracer fraction exceeds 10\%.}
        \label{fig:simulations_d}
    \end{subfigure}

    \caption{RHD jet simulation snapshots at three evolutionary times: 
    8.16~kyr, 24.48~kyr, and 40.79~kyr (from left to right in each panel). Panels \subref{fig:simulations_a}--\subref{fig:simulations_d} show, respectively, the tracer fraction, Lorentz factor, density, and thermal pressure. The simulation is performed in a Cartesian domain $(x, y, z)$ spanning $[-17.5, 17.5] \times [-17.5, 17.5] \times 
    [2.5, 150]$~pc.}
    \label{fig:simulations}
\end{figure*}

Figure~\ref{fig:simulations} shows the simulated jet at three 
evolutionary times, 8.16~kyr, 24.48~kyr, and 40.79~kyr, 
illustrating the spatial distribution of the tracer fraction, 
Lorentz factor, rest-mass density, and thermal pressure. The 
simulation is initialized with the parameters listed in 
Table~\ref{tab:setup_simulation}, chosen to reproduce the 
parsec-scale morphology and moderately relativistic velocities 
observed for LEDA~55267 and LEDA~58287~\citep{Baldi_2019b, 
Giovannini_2023}, following the recollimation shock instability 
framework of~\citet{Costa_2024}.

The tracer distribution in Figure~\ref{fig:simulations_a} reveals that the jet material is rapidly disrupted as the flow evolves. At the earliest snapshot (8.16~kyr), the jet propagates in a relatively well-collimated manner within the first few parsecs, but already shows signs of lateral spreading and mixing at its head. By 24.48~kyr, the jet has lost its coherent columnar structure and the tracer is distributed across a broad, turbulent volume, indicating that the jet has become heavily stratified. At 40.79~kyr, the tracer fills an even wider region and no clear jet boundary can be identified, confirming that the flow has been effectively disrupted well before reaching scales of several tens of parsecs. This behavior is consistent with the results of~\citet{Costa_2024} and~\citet{Costa_2026}, who report qualitatively similar disruption in RHD simulations of light jets, and with~\citet{Borodina_2025}, who show that jets propagating through a turbulent interstellar medium are similarly confined to compact scales.

The physical mechanism driving this disruption is visible in the sequence of panels. As soon as the jet begins to expand beyond the launching region, it becomes under-pressured relative to the external medium, triggering the formation of a recollimation shock. This shock breaks the jet axisymmetry and generates centrifugal instabilities at the jet–ambient medium interface, driving the flow into a highly turbulent regime. Subsequent recollimation events are suppressed by the growing instabilities, preventing any sustained re-acceleration of the flow.

The Lorentz factor maps in~Figure~\ref{fig:simulations_b} quantify the rapid deceleration that follows. At 8.16~kyr, relativistic velocities with $\Gamma \gtrsim 2$ are confined to the innermost few parsecs near the jet base. By 24.48~kyr, the high-$\Gamma$ region has shrunk further and the bulk of the jet material is already sub-relativistic, with $\Gamma \sim 1$. At 40.79~kyr, essentially no relativistic motion is detectable beyond the immediate launch zone. This deceleration profile is broadly consistent with the moderately relativistic jet velocities inferred observationally for FR0 sources~\citep{Baldi_2019b, Giovannini_2023} and with the kinematic constraints discussed by~\citet{Boula_2025} in the RMHD context.

The density maps in Figure~\ref{fig:simulations_c} illustrate how the ambient medium is progressively entrained into the jet. At early times, the high-density external gas remains largely confined to the region surrounding the jet cone. As the instabilities develop, the density contrast between the jet and the ambient medium decreases substantially, reflecting efficient mass loading and mixing. By 40.79~kyr, the density distribution within the tracer-defined jet volume approaches that of the surrounding medium, signaling that the jet has been effectively decelerated and disrupted through momentum exchange with the environment. This mixing process is qualitatively similar to the entrainment reported by~\citet{Borodina_2025} for jets propagating through a clumpy ISM, and reinforces the picture in which jet disruption in FR0s is driven primarily by interaction with the ambient medium rather than by intrinsic properties of the central engine alone.

The pressure evolution shown in Figure~\ref{fig:simulations_d} completes this picture. At 8.16~kyr, the jet head maintains a pressure significantly above that of the surrounding medium, consistent with a ram-pressure dominated propagation phase. However, as the recollimation instabilities develop, the jet pressure drops sharply and approaches the ambient value of $p_{\rm ext} = 1.5 \times 10^{-9}$~dyne\,cm$^{-2}$ throughout most of the flow volume. The pressure equilibration is essentially complete by 40.79~kyr, at which point the jet has lost its ability to drive further expansion. Taken together, the four panels of Figure~\ref{fig:simulations} consistently show that the simulated jet fails to propagate beyond a few tens of parsecs, in good agreement with the compact radio morphologies observed for LEDA~55267 and LEDA~58287 \citep{Giovannini_2023} and with the general picture of FR0 jet confinement discussed in the recent 
literature~\citep{Costa_2024, Costa_2026, Boula_2025, Borodina_2025}. The thermal pressures extracted from the simulated jet, particularly the values reached at late evolutionary times 
when the flow approaches pressure equilibrium with the ambient 
medium, will be used in Section~\ref{sec:beta_p} to estimate 
the magnetic field strength via the plasma $\beta_{\rm p}$ 
parameter, independently of the SED fits, providing a 
consistency check between the dynamical and radiative analyses.

\begin{figure}
    \centering
    \includegraphics[width=0.99\columnwidth]{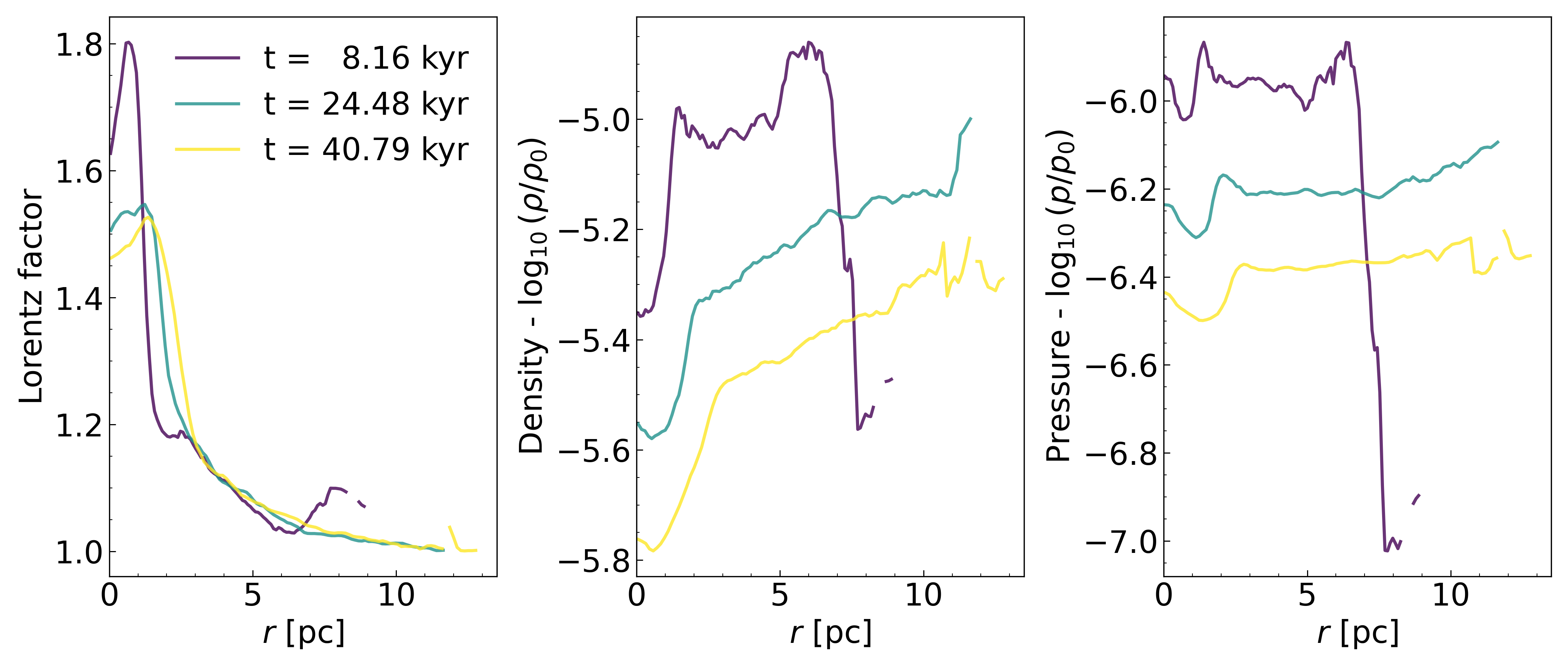}
    \caption{Azimuthally averaged radial profiles of the Lorentz factor
    (left), rest-mass density (middle), and thermal pressure (right) at
    three evolutionary times: $t = 8.16$~kyr, $24.48$~kyr, and $40.79$~kyr. Each profile is obtained by averaging the corresponding quantity over cylindrical shells in the plane transverse to the jet axis, restricted to cells where the tracer fraction exceeds $10\%$, thereby isolating the jet material from the ambient medium. The profiles illustrate the rapid radial decline of the Lorentz factor, the progressive density flattening driven by entrainment, and the tendency of the pressure profile to relax towards the external value as the jet evolves.}
    \label{fig:radial_profiles}
\end{figure}

To complement the three-dimensional view shown above, azimuthally averaged characterization of the simulated jet, Figure~\ref{fig:radial_profiles} shows the radial profiles of the Lorentz factor, density, and pressure at the same three evolutionary times. The profiles are obtained by averaging each quantity within cylindrical shells in the transverse plane, considering only regions where the tracer fraction exceeds $10\%$, so that they describe the jet material rather than the ambient medium. The Lorentz factor decreases sharply with radius at all epochs, confirming that relativistic velocities are confined to the innermost regions and rapidly become sub-relativistic elsewhere, in agreement with the spatial maps and consistent with the turbulent mixing scenario. The density profiles increase gradually with radius, reflecting the entrainment of ambient gas, an effect that becomes progressively more pronounced as time advances and is most evident at $40.79$~kyr, in line with the diffuse structure visible in Figure~\ref{fig:simulations_a} at this epoch. The pressure profiles flatten radially as the system evolves, approaching the external value at large radii, which signals the transition from a driven jet phase to a disturbed flow dominated by turbulence. Overall, the radial profiles reinforce the conclusions drawn from the three-dimensional maps regarding the rapid deceleration, efficient mixing, and early disruption of the flow.

\subsection{Synthetic morphology and emission maps}
\label{sec:morphology}

\begin{figure}
    \centering
    \includegraphics[width=0.95\columnwidth]{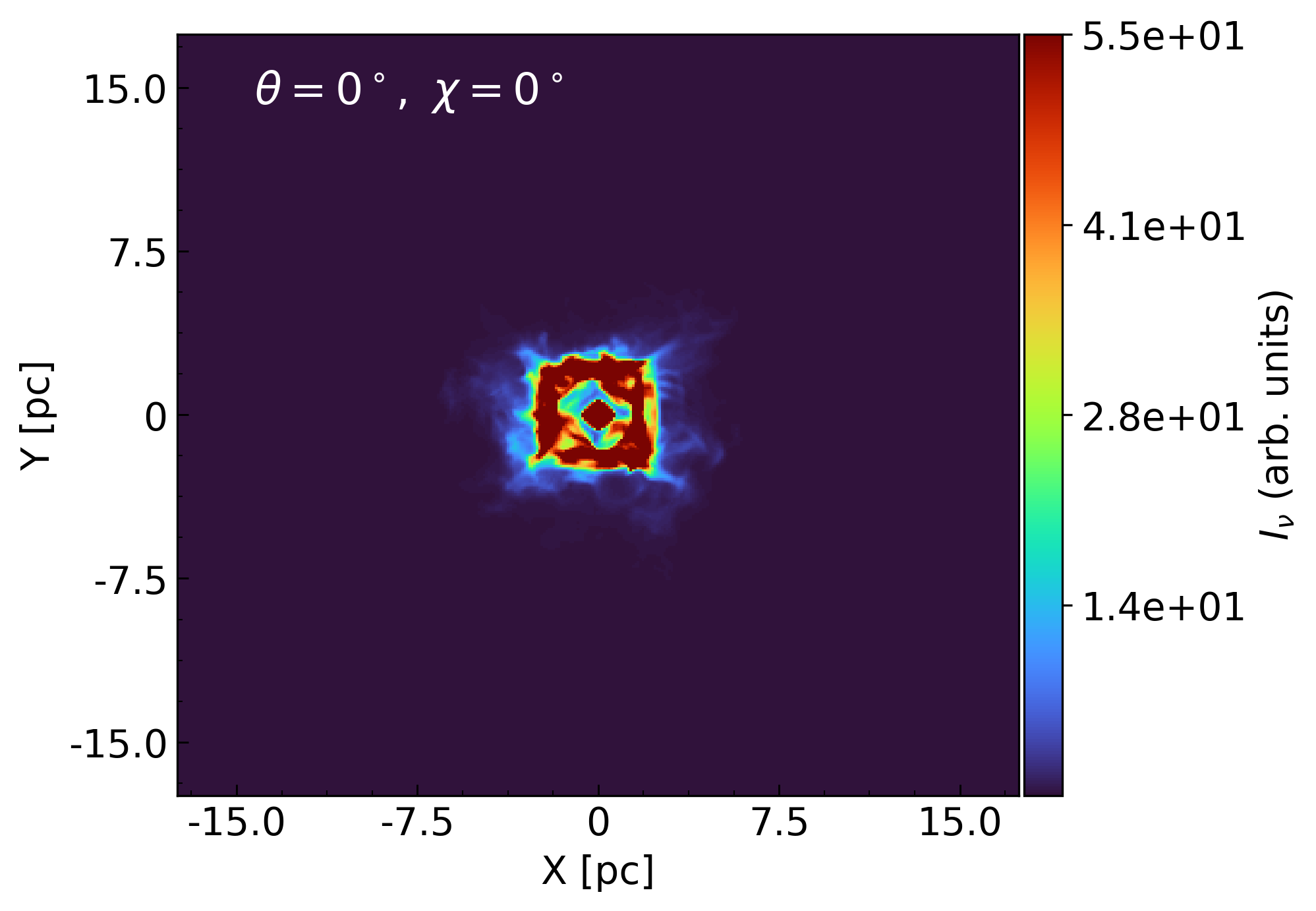} \\
    \includegraphics[width=0.95\columnwidth]{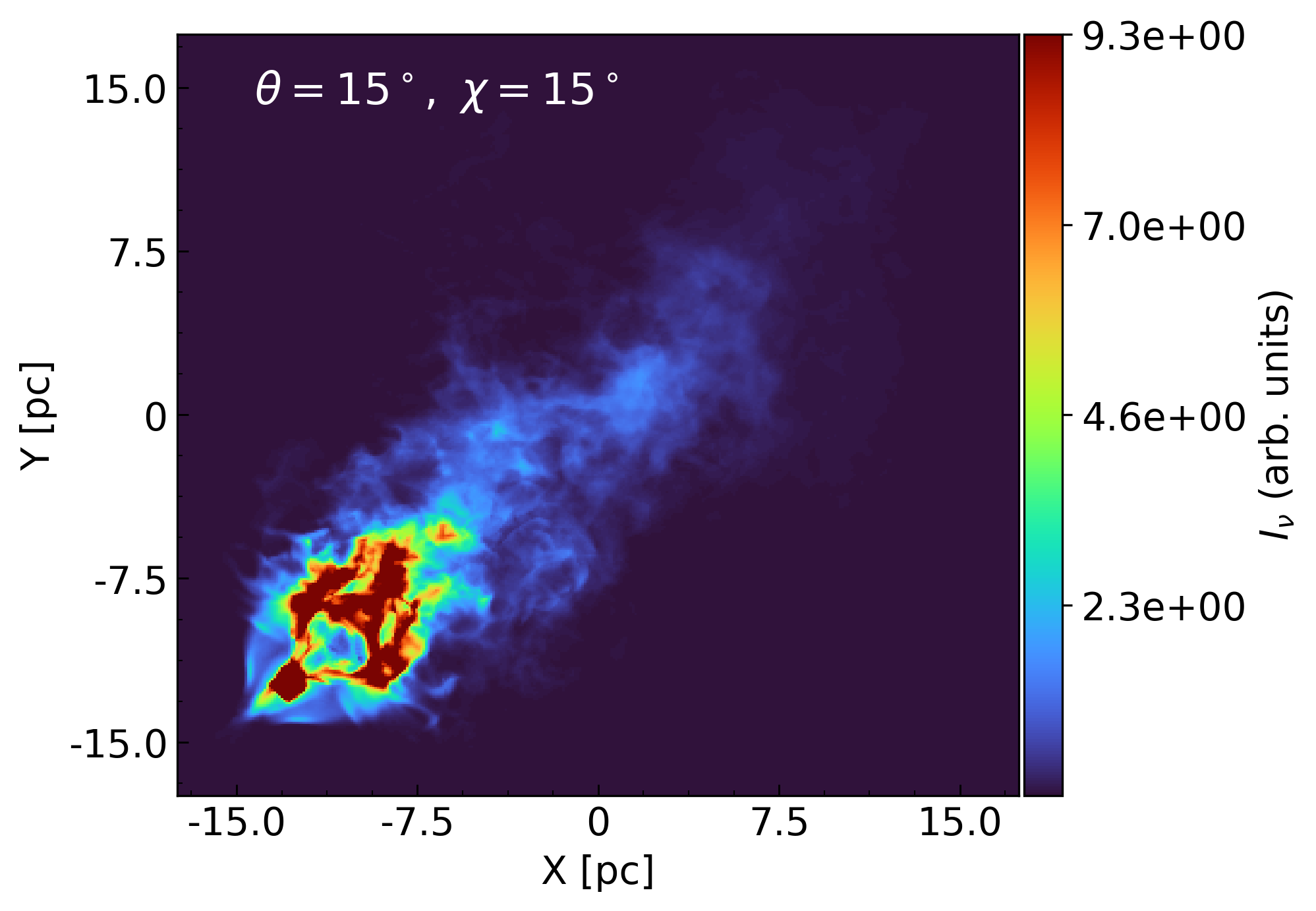}
    \caption{Synthetic synchrotron emission maps of the simulated jet projected onto the $(x, y)$ plane. The upper panel represents LEDA~55267, obtained without jet reorientation ($\theta = 0^{\circ}$, $\chi = 0^{\circ}$), corresponding to a configuration approximately aligned with the line of sight. The lower panel represents LEDA~58287, obtained after rotating the jet by~$\theta = 15^{\circ}$ around the $x$ axis and~$\chi = 15^{\circ}$ around the $y$ axis, producing a more asymmetric projected morphology consistent with the observed radio structure. The color bar indicates the synchrotron intensity in arbitrary units.}
    \label{fig:mapas_sinteticos}
\end{figure}

To evaluate the morphology and structure of the simulated jets 
and compare them with the radio maps presented in 
Figure~\ref{fig:radio_ledas}, we produce synthetic maps of the 
synchrotron emission based on the jet simulations. These maps 
serve a dual purpose: they allow a direct morphological 
comparison with the observed radio structures, and they identify the regions of enhanced emissivity that will inform the SED 
modeling presented in Section~\ref{sec:fit_lept_hadr}. The 
magnetic field in the simulated jet is assumed to be 
proportional to the thermal pressure, an approximation widely 
adopted in the literature when direct observational constraints 
on the field geometry are unavailable~\citep{Gomez_1995, Nawaz_2016}. The physical justification for this assumption and its implications for the magnetization of the emitting regions are discussed in Section~\ref{sec:beta_p}. Under this prescription, the synchrotron emissivity in the rest frame is computed 
following~\citet{Sutherland_2007} as
\begin{equation}
    j_{\nu} = \lambda\, p^{\alpha + 3/2}\, \delta^{2 + \alpha},
    \label{eq:emissivity}
\end{equation}
where $\lambda$ is a parameter that isolates the jet plasma from 
the ambient medium, $p$ is the gas pressure assumed proportional 
to the total particle pressure, and the spectral index is fixed 
at $\alpha = 1.0$. The Doppler factor $\delta$ accounts for 
relativistic beaming effects and is computed locally in each 
cell as
\begin{equation}
    \delta = \frac{1}{\Gamma(1 - \beta\cos\theta')},
    \label{eq:Doppler_factor}
\end{equation}
\noindent where $\beta$ is the flow velocity in units of $c$, 
$\Gamma$ is the Lorentz factor, and $\theta' = 
\cos^{-1}(v'_{\rm LOS}/|\mathbf{v}'|)$ is the angle between 
the rotated velocity vector $\mathbf{v}'$ of each cell and the 
line-of-sight (LOS) direction, with $v'_{\rm LOS}$ being the 
LOS component of $\mathbf{v}'$~\citep{Nawaz_2016}. The 
synthetic emissivity along the line of sight is then obtained 
by integrating the rotated three-dimensional emissivity field as
\begin{equation}
    I_{\nu} = \int j_{\nu}\, \mathrm{d}s.
    \label{eq:synthetic_emissivity}
\end{equation}
As discussed in~\citet{Nawaz_2016}, the projected morphology 
of the observed source depends on both the inclination angle 
between the jet propagation axis and the line of sight, and 
the azimuthal viewing angle. For LEDA~55267, we project the 
emissivity integrated along the jet evolution axis $z$ onto 
the $(x, y)$ plane without applying any reorientation, 
corresponding to a configuration in which the simulated jet 
is approximately aligned with the line of sight, producing a 
compact, centrally concentrated emission pattern consistent 
with the unresolved radio morphology of 
LEDA~55267~\citep{Giovannini_2023}. For LEDA~58287, we apply a three-dimensional reorientation to the simulated jet volume. Defining the $z$ axis as the initial line of sight, the jet is rotated through two successive transformations: a rotation of $\theta = 15^{\circ}$ in the $(y,z)$ plane around the $x$ axis, which inclines the jet axis with respect to the plane of the sky, followed by a rotation of $\chi = 15^{\circ}$ in the $(x,z)$ plane around the $y$ axis, which adjusts the position angle of the jet in the two-dimensional projection. This produces a more asymmetric projected
distribution of the synchrotron emission, in good agreement with the
rotated outer structure observed in LEDA~58287 \citep{Giovannini_2023}. The resulting synthetic synchrotron emission maps are presented in 
Figure~\ref{fig:mapas_sinteticos}.

The viewing angles adopted here are not uniquely determined 
and should be understood as effective geometrical 
configurations capable of reproducing the observed source 
morphologies, rather than as definitive measurements of the 
true jet orientations. This is consistent with the findings 
of~\citet{Giovannini_2023}, who show that moderately relativistic FR0 jets are compatible with a wide range of inclination angles with respect to the line of sight. Furthermore, \citet{Massaro_2020} showed that BL~Lac objects
inhabit environments statistically consistent with those of FR0 radio
galaxies, suggesting that a fraction of BL~Lacs may correspond to FR0
sources observed at small viewing angles. Within this broader orientation framework, the aligned configuration adopted for LEDA~55267 should be interpreted as a limiting case rather than as a definitive measurement of the jet inclination.

A grid-like pattern is visible at the base of the jet in 
Figure~\ref{fig:mapas_sinteticos}. This is a numerical 
artifact arising from the relatively small number of 
computational cells across the jet cross-section at the 
injection boundary, which prevents the base of the jet from 
appearing fully cylindrical in projection. The structure is therefore a resolution effect with no physical significance, 
and is expected to smooth out at higher numerical resolution. 
It does not affect the morphological features of the scales 
of interest, namely the overall extent and asymmetry of the 
projected emission distribution.

\begin{figure}
    \centering
    \includegraphics[width=0.95\columnwidth]{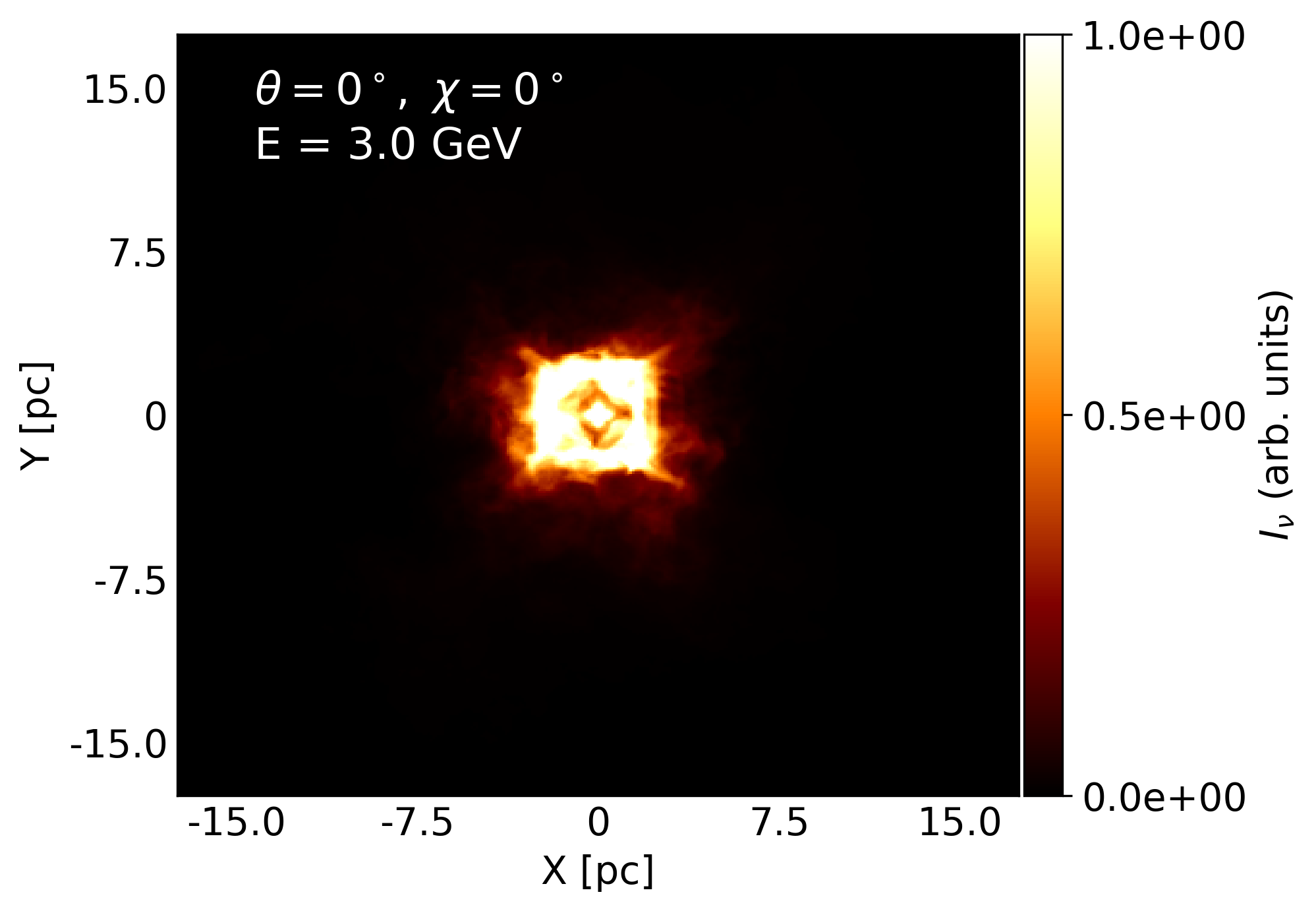} \\
    \includegraphics[width=0.95\columnwidth]{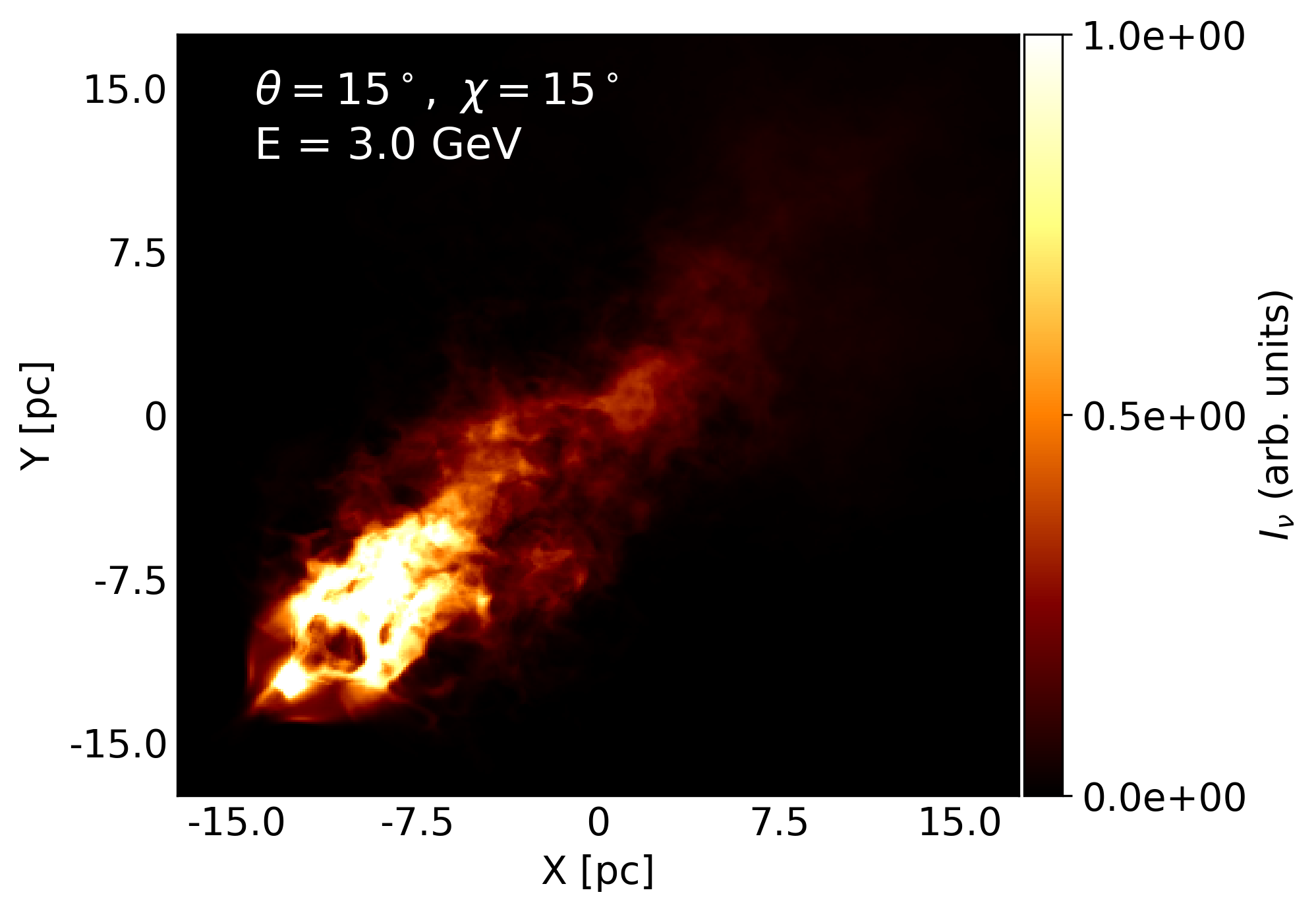}
    \caption{Synthetic inverse Compton emission maps of the simulated jet projected onto the $(x, y)$ plane, computed at an energy of 3.0~GeV. The upper panel represents LEDA~55267, obtained without jet reorientation ($\theta = 0^{\circ}$, $\chi = 0^{\circ}$). The lower panel represents LEDA~58287, obtained after rotating the jet by~$\theta = 15^{\circ}$ around the $x$ axis and~$\chi = 15^{\circ}$ around the $y$ axis. Emission is restricted to regions where the local Mach number exceeds unity, for which an electron acceleration efficiency of $f_{\rm e} = 0.5$ is assumed. The color bar indicates the inverse Compton intensity in arbitrary units.}
    \label{fig:mapas_sinteticos_ic}
\end{figure}

For completeness, we also produce synthetic maps of the 
inverse Compton emission at an energy of 3.0~GeV, using the 
same angular orientations as in Figure~\ref{fig:mapas_sinteticos}. These maps are computed with the publicly available code 
\texttt{NAIMA}\footnote{Additional information, including 
installation and implementation details, is available at 
\href{http://naima.readthedocs.org}{http://naima.readthedocs.org}.}~\citep{naima_Zabalza_2015}, 
adopting a power-law electron energy distribution with an 
exponential cutoff whose spectral index and cutoff energy are 
taken from the leptonic fits described in Section~\ref{sec:fit_lept_hadr}. Particle acceleration is 
assumed to operate only in regions where the local Mach number 
exceeds unity, with an electron acceleration efficiency 
$f_{\rm e} = 0.5$. The resulting synthetic inverse Compton 
emission maps are presented in 
Figure~\ref{fig:mapas_sinteticos_ic}.

Despite originating from distinct radiative mechanisms, the
synthetic synchrotron and inverse Compton maps share similar
morphological features. This is the expected outcome of
our framework, since the inverse Compton emission is primarily
driven by synchrotron self-Compton (SSC) processes arising from
the same population of relativistic electrons responsible for the
synchrotron radiation. In both cases, the emission is concentrated
in the central region of the jet, where the tracer analysis of
Section~\ref{sec:simulation_results} identified the most active
zones of jet--ambient medium interaction, associated with the
recollimation shocks and turbulent mixing at the jet boundary.
The co-spatiality of the maximum emission intensities
does not simply reflect the radiative coupling between synchrotron and inverse Compton processes, but also underscores the role of these dynamical structures as common sites of particle acceleration.
This result provides independent support for the assumption of co-spatial
emitting populations adopted in the SED modeling of
Section~\ref{sec:fit_lept_hadr}. The synthetic maps reproduce the
main morphological features of the observed radio emission in
Figure~\ref{fig:radio_ledas} reasonably well, lending physical
credibility to the overall simulation setup.

\subsection{Leptonic and lepto-hadronic modeling}
\label{sec:fit_lept_hadr}

In this section, we present the SED fitting results for LEDA~55267 and LEDA~58287. Their multi-wavelength luminosities, together with the corresponding telescopes/instruments, are summarized in Table~\ref{tab:Multi-wavelength-luminosities}. For the currently available flux data points, we consider only the leptonic scenario, following the approach  of~\citet{Khatiya_2024}. When the simulated CTAO flux points are included, we explore both leptonic and lepto-hadronic scenarios 
to assess whether a hadronic contribution may play a role in the 
production of the highest-energy gamma-ray emission. We stress 
from the outset, that the preference for the lepto-hadronic 
scenario discussed below is a theoretical prediction based on 
simulated CTAO observations, not a constraint imposed by 
currently available data. The simulated CTAO flux points are 
generated under the assumptions described in 
Section~\ref{sec:ctao}, namely a 200~hour observation with 
the CTAO North array at a zenith angle of $20^{\circ}$, using 
the spectral models fitted to the existing multiwavelength data 
as input and applying EBL attenuation following 
\citet{Saldana-Lopez_2021}. These points, therefore, represent the flux values that would be measured if the source emission 
followed the input spectral model, and their role is to extend 
the observational coverage into the TeV regime where leptonic 
and lepto-hadronic scenarios diverge. The best-fit parameters 
obtained with \texttt{NAIMA} for each source and scenario are 
listed in Table~\ref{tab:parameters-sed}.

\begin{table}
    \centering
    \setlength\tabcolsep{0.2cm}
    \caption{Representative multi-wavelength luminosities of LEDA~55267 and LEDA~58287, together with the telescopes/instruments from which the fluxes are taken.}
    \label{tab:Multi-wavelength-luminosities}
    \begin{tabular}{lccc}
        \hline\hline
        Band & Energy & Luminosity [erg s$^{-1}$] & Telescope \\
        \hline
        \multicolumn{4}{c}{\textit{LEDA~55267}} \\
        \hline
        Radio        & 1.4 GHz   & $6.58 \times 10^{38}$ & NVSS/FIRST \\
        X-ray        & 1.4 keV   & $2.47 \times 10^{41}$ & Swift \\
        $\gamma$-ray & 5.5 GeV   & $7.08 \times 10^{41}$ & Fermi \\
        $\gamma$-ray & 55 GeV    & $4.18 \times 10^{41}$ & Fermi \\
        \hline
        \multicolumn{4}{c}{\textit{LEDA~58287}} \\
        \hline
        Radio        & 1.4 GHz   & $1.25 \times 10^{39}$ & NVSS/FIRST \\
        X-ray        & 0.5-7 keV & $1.49 \times 10^{41}$ & Chandra \\
        $\gamma$-ray & 5.5 GeV   & $5.84 \times 10^{41}$ & Fermi \\
        $\gamma$-ray & 55 GeV    & $4.64 \times 10^{41}$ & Fermi \\
        \hline\hline
    \end{tabular}
\end{table}

\begin{figure*}
    \centering
    \includegraphics[width=1.99\columnwidth]{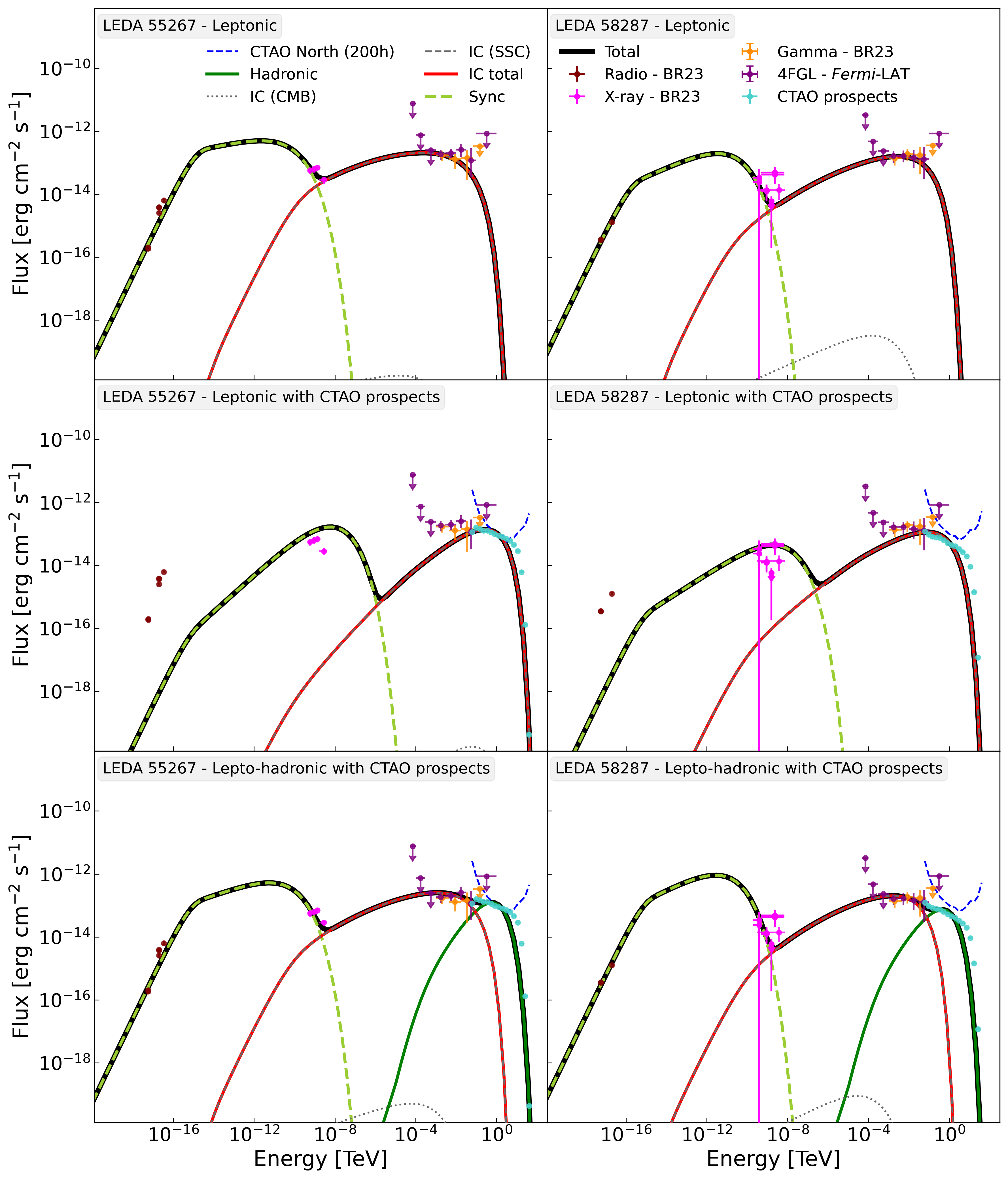}
    \caption{Broadband SED models for LEDA~55267 (left panels) and LEDA~58287 (right panels). From top to bottom, the panels show the leptonic model, the leptonic model with CTAO prospects, and the lepto-hadronic model with CTAO prospects. Data points: brown, magenta, and orange correspond to radio, X-ray, and gamma-ray measurements from~\citetalias{Boughelilba_2023}, respectively; purple to gamma-ray flux points from the 4FGL \textit{Fermi}-LAT catalog~\citep{Abdollahi_2020, Ballet_2023}; and light-blue to simulated CTAO flux points. Curves: blue dashed, CTAO sensitivity; green solid, hadronic flux contribution; gray dotted, inverse Compton emission from the CMB; gray dashed, synchrotron self-Compton emission from the emitting region; red solid, total inverse Compton emission; olive dashed, synchrotron emission; black solid, total emission.}
    \label{fig:seds}
\end{figure*}

For both sources, the leptonic best-fit parameters are in good 
agreement with those reported by \citet{Khatiya_2024}, with 
magnetic field strengths in the range $0.002$--$0.02$~G, 
which are also consistent with the estimates of~\citet{Merten_2021}. The small differences with respect to~\citet{Khatiya_2024} are attributable to the inclusion of the most recent 4FGL \textit{Fermi}-LAT flux points~\citep{Abdollahi_2020, Ballet_2023}. As visible in the top panels of Figure~\ref{fig:seds}, the leptonic model reproduces the observed radio, X-ray, and GeV data satisfactorily. The 
synchrotron component peaks in the radio band, while the inverse Compton component, dominated by SSC emission, accounts 
for the GeV flux. The negligible contribution from CMB inverse 
Compton scattering is consistent with the compact nature of FR0 
jets, in which the internal synchrotron radiation field greatly 
exceeds any external seed photon field, as also noted 
by~\citet{Khatiya_2024} and~\citet{Boughelilba_2023}. The higher magnetic field strengths inferred here, when compared to those typically reported for extended FRI jets, follow naturally from the compact nature of FR0 sources, since the SED is being modeled close to the emitting region, where the magnetic energy density inherited from the launching site has not yet been significantly dissipated.

However, the leptonic model alone does not constrain the TeV 
regime, since \textit{Fermi}-LAT observations do not extend to 
these energies. The middle panels of Figure~\ref{fig:seds} show 
that, when the simulated CTAO flux points are included, the 
leptonic model struggles to reproduce the TeV emission 
simultaneously with the GeV data, requiring a harder electron 
spectrum and a significantly higher cutoff energy ($E_{\rm 
cut,e} \sim 3$~TeV for LEDA~55267 and $\sim 2$~TeV for 
LEDA~58287) compared to the fits based on currently available 
data. This tension motivates the exploration of a lepto-hadronic 
scenario, in which relativistic protons interact with ambient 
photons or matter to produce additional high-energy emission 
through pion decay~\citep{Dermer_2009, Tavecchio_2018}. In the 
lepto-hadronic fits shown in the bottom panels of 
Figure~\ref{fig:seds}, the hadronic component rises steeply at 
TeV energies and accounts naturally for the simulated CTAO flux 
points, while the leptonic component continues to describe the 
radio and GeV data.

The hadronic model parameters listed in 
Table~\ref{tab:parameters-sed} deserve brief discussion in terms 
of their physical implications. The proton spectral indices 
$p_{\rm p} = 0.01$ for both sources are unusually hard, 
suggesting that the proton energy distribution is nearly flat 
at injection, which may reflect an efficient diffusive shock 
acceleration at the recollimation shocks identified in our RHD 
simulations~\citep{Blandford_1987, Kowal_2011}. The proton 
cutoff energies $E_{\rm cut,p} = 3.78$~TeV for LEDA~55267 and 
$E_{\rm cut,p} = 2.86$~TeV for LEDA~58287 are consistent with 
maximum energies achievable in compact, strongly magnetized 
regions of parsec-scale jets, where the Larmor radius of 
accelerated protons remains smaller than the size of the 
emitting region. The electron-to-proton energy ratios $K_{\rm 
ep} = 0.44$ for LEDA~55267 and $K_{\rm ep} = 0.54$ for 
LEDA~58287 indicate that protons carry a substantial fraction 
of the total particle energy, with the total non-thermal energy 
budget implied by these fits remaining within the range expected 
for FR0 jets given their radio luminosities of $10^{38.69}$ and 
$10^{39.15}$~erg~s$^{-1}$~\citep{Baldi_2018}, and consistent 
with the jet powers inferred from the simulation setup described 
in Section~\ref{sec:jet_parameters}. This suggests that the 
lepto-hadronic scenario does not require energetically extreme 
conditions and is physically compatible with typical FR0 
systems.

The recollimation shocks identified in our RHD simulations, 
discussed in Section~\ref{sec:simulation_results}, are 
efficient particle acceleration sites where both electrons 
and protons can be energized via diffusive shock 
acceleration~\citep{Blandford_1987, Kowal_2011, 
Dal_Pino_2015, Kagan_2015}. The compact, parsec-scale 
emitting regions inferred from the SED fits are consistent 
with the strongly magnetized conditions required for proton 
confinement and acceleration to TeV--PeV 
energies~\citep{Boula_2025}. Moreover, FR0 sources are the 
most numerous jetted AGN population in the local 
Universe~\citep{Baldi_2018}, and even a modest individual 
contribution to the diffuse cosmic-ray and neutrino 
backgrounds could be collectively 
significant~\citep{Tavecchio_2018, Merten_2021, 
Lundquist_2025}.

To determine which model provides the best statistical 
description of the data, we adopt the Bayesian Information 
Criterion~\citep[BIC;][]{Kass_1995, Szydlowski_2015, 
Trotta_2008}. The evidence in favor of model $i$ over model $j$ 
is quantified through the log Bayes factor $|\ln B_{ij}| = 
-({\rm BIC}_i - {\rm BIC}_j)/2$, interpreted according to the 
following scale: inconclusive if $|\ln B_{ij}| \leq 1.0$, weak 
evidence if $1.0 < |\ln B_{ij}| \leq 2.5$, moderate evidence 
if $2.5 < |\ln B_{ij}| \leq 5.0$, and strong evidence if 
$|\ln B_{ij}| > 5.0$. For LEDA~55267, the leptonic and 
lepto-hadronic fits including CTAO prospects yield ${\rm BIC} = 
5691.13$ and ${\rm BIC} = 4814.84$, respectively, giving 
$|\ln B_{ij}| = 438.15$. For LEDA~58287, the corresponding 
values are ${\rm BIC} = 3439.98$ and ${\rm BIC} = 2506.30$, 
giving $|\ln B_{ij}| = 466.84$. In both cases, the evidence 
strongly favors the lepto-hadronic scenario when the simulated 
CTAO prospects are included. This conclusion is further 
supported by the model weights proposed 
by~\citet{Burnham_2002}. Defining $\Delta_i = {\rm BIC}_i - 
{\rm BIC}_{\rm min}$, the weight of model $i$ is
\begin{equation}
    w_i = \frac{\exp(-\frac{1}{2}\Delta_i)}{\sum_{r=1}^{R} 
    \exp(-\frac{1}{2}\Delta_r)},
    \label{eq:bic_weight}
\end{equation}
which assigns a probability of 100\% to the lepto-hadronic 
model for both sources when the CTAO prospects are incorporated.

We stress once more that this preference does not arise from 
the currently available data, but it emerges only when the 
simulated TeV flux points are included, and should therefore be 
interpreted as a theoretical prediction to be tested by future 
CTAO observations. At GeV energies, leptonic and lepto-hadronic 
models produce degenerate predictions, since the hadronic 
component contributes negligibly below $\sim 1$~TeV. It is 
precisely in the TeV regime that the two scenarios diverge, 
with the hadronic pion-decay emission rising steeply and 
dominating the total flux, as clearly visible in the bottom 
panels of Figure~\ref{fig:seds}. This degeneracy at GeV 
energies and divergence at TeV energies is a well-known 
challenge in AGN jet 
modeling~\citep{Bottcher_2013} and underscores the unique 
diagnostic power of the CTAO. Future observations of approximately 200~hours with the CTAO 
North array at a zenith angle of $20^{\circ}$ will be essential 
to discriminate between these scenarios and to establish whether 
FR0 radio galaxies are genuine sites of hadronic particle 
acceleration and significant contributors to the diffuse 
high-energy neutrino and cosmic-ray backgrounds. We note that this observation time represents a conservative reference rather 
than a sufficient threshold, since as shown in 
Figure~\ref{fig:seds}, 200~hours are not enough to detect 
significant flux points in the leptonic scenario nor to fully 
resolve the lepto-hadronic component at the lowest TeV energies, 
reflecting the intrinsic faintness of FR0 sources compared to 
the classical AGN targets of Cherenkov observatories. This 
reference nevertheless illustrates the potential contribution 
of the CTAO to FR0 science and motivates future studies of how 
detection significance scales with observation time for sources 
of this luminosity class.

\begin{table}
    \centering
    \setlength\tabcolsep{0.12cm}
    \caption{Best-fit parameters of the leptonic, leptonic with CTAO 
    prospects, and lepto-hadronic with CTAO prospects models for 
    LEDA~55267 and LEDA~58287. Dashes indicate parameters not applicable to the corresponding scenario. $K_{\rm ep}$ is the ratio of the total energy in relativistic electrons to that in relativistic protons.}
    \label{tab:parameters-sed}
    \begin{tabular}{lccc}
        \hline\hline
        Parameter & Leptonic & Leptonic$+$CTAO & Lepto-hadronic$+$CTAO \\
        \hline
        \multicolumn{4}{c}{\textit{LEDA~55267}} \\
        \hline
        $p_{\rm e}$              & 2.72                  & 1.84                  & 2.55                  \\
        $E_{\rm cut,e}$ [TeV]   & 0.21                  & 3.12                  & 0.25                  \\
        $B$ [G]                  & $1.75 \times 10^{-2}$ & $1.00 \times 10^{-2}$ & $1.06 \times 10^{-2}$ \\
        $R$ [cm]                 & $3.51 \times 10^{18}$ & $1.00 \times 10^{17}$ & $4.46 \times 10^{18}$ \\
        $p_{\rm p}$              & ---                   & ---                   & 0.01                  \\
        $E_{\rm cut,p}$ [TeV]   & ---                   & ---                   & 3.78                  \\
        $K_{\rm ep}$             & ---                   & ---                   & 0.44                  \\
        \hline
        \multicolumn{4}{c}{\textit{LEDA~58287}} \\
        \hline
        $p_{\rm e}$              & 2.42                  & 2.17                  & 2.34                  \\
        $E_{\rm cut,e}$ [TeV]   & 0.31                  & 2.56                  & 0.12                  \\
        $B$ [G]                  & $2.56 \times 10^{-3}$ & $1.00 \times 10^{-2}$ & $1.06 \times 10^{-2}$ \\
        $R$ [cm]                 & $8.25 \times 10^{18}$ & $1.00 \times 10^{17}$ & $8.60 \times 10^{18}$ \\
        $p_{\rm p}$              & ---                   & ---                   & 0.01                  \\
        $E_{\rm cut,p}$ [TeV]   & ---                   & ---                   & 2.86                  \\
        $K_{\rm ep}$             & ---                   & ---                   & 0.54                  \\
        \hline\hline
    \end{tabular}
\end{table}

\subsection{Plasma magnetization and $\beta_{\rm p}$ estimates}
\label{sec:beta_p}

\begin{figure}
    \centering
    \includegraphics[width=0.95\columnwidth]{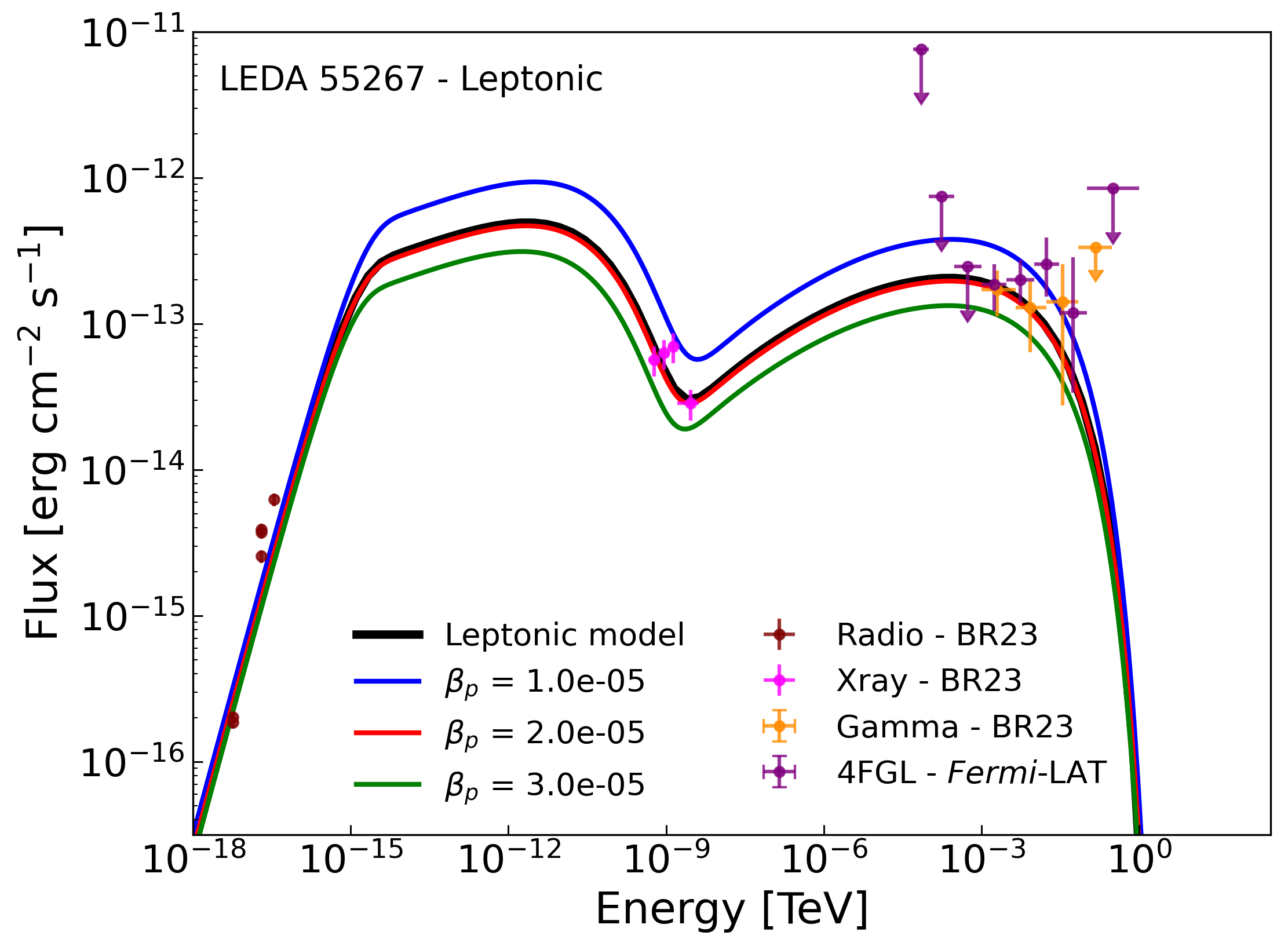} \\
    \includegraphics[width=0.95\columnwidth]{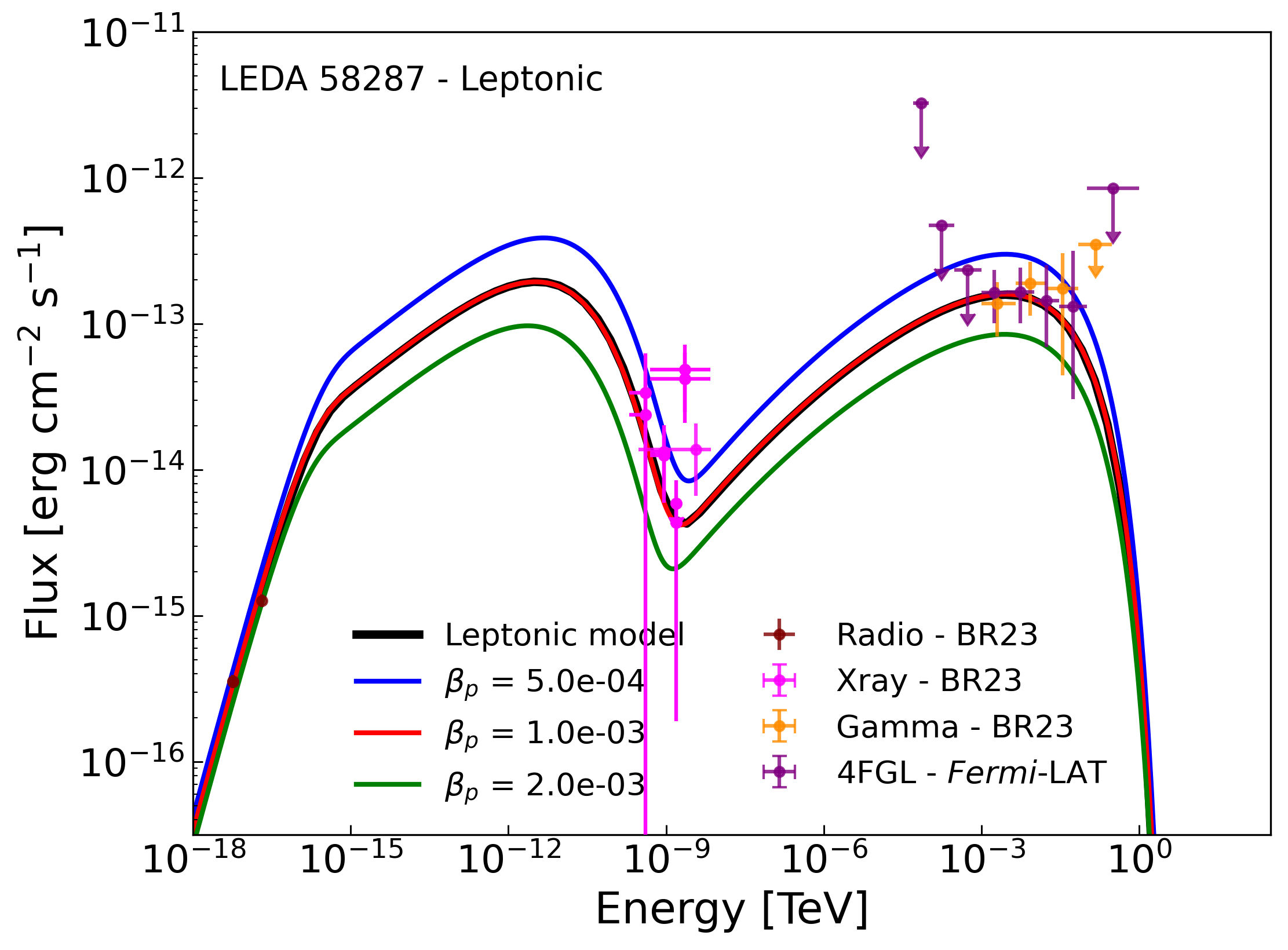}
    \caption{Leptonic model fits for LEDA~55267 (upper panel) and LEDA~58287 (lower panel) for different values of the plasma parameter $\beta_{\rm p}$. In each panel, the black curve corresponds to the reference leptonic model with the parameters listed in Table~\ref{tab:parameters-sed}, while the blue, red, and green curves show the fits obtained for varying $\beta_{\rm p}$. For LEDA~55267, the three curves correspond to $\beta_{\rm p} = 1.0 \times 10^{-5}$, $2.0 \times 10^{-5}$, and $3.0 \times 10^{-5}$, respectively, with the red curve ($\beta_{\rm p} = 2.0 \times 10^{-5}$) providing the best agreement with the data. For LEDA~58287, the curves correspond to $\beta_{\rm p} = 4.0 \times 10^{-4}$, $1.0 \times 10^{-3}$, and $2.0 \times 10^{-3}$, respectively, with the red curve ($\beta_{\rm p} = 1.0 \times 10^{-3}$) providing the best agreement.}
    \label{fig:mag_prs_ledas}
\end{figure}

The synthetic emission maps presented in Section~\ref{sec:morphology} identified the central regions of the jet, where recollimation shocks and turbulent mixing are most active, as the primary sites of synchrotron and inverse Compton emission. To assess the consistency between the jet simulations and the leptonic emission scenario, we estimate the magnetic field strength by relating it to the thermal pressure of the simulated flow. Although the magnetic field is not evolved self-consistently in the RHD simulations, this approach provides an order-of-magnitude estimate widely adopted in the literature when direct observational constraints on the field geometry 
are unavailable~\citep{Wilson_1983, Gomez_1995, Seo_2023, 
de_Oliveira_2025}. Under the assumption that the simulation pressure corresponds to the thermal pressure, the magnetic field strength is estimated through the plasma beta parameter
\begin{equation}
    \beta_{\rm p} = \frac{P}{P_{\rm B}} = \frac{8\pi P}{B^{2}},
\end{equation}
\noindent where $P_{\rm B} = B^{2}/8\pi$ is the magnetic pressure, so that
\begin{equation}
    B = \sqrt{\frac{8\pi P}{\beta_{\rm p}}}.
    \label{eq:B_beta}
\end{equation}
We perform a leptonic analysis using \texttt{NAIMA}~\citep{naima_Zabalza_2015}, adopting the best-fit 
parameters from Table~\ref{tab:parameters-sed} for all quantities 
except the magnetic field, which is determined by Equation~\ref{eq:B_beta} for a range of $\beta_{\rm p}$ values. The value of $\beta_{\rm p}$ that best reproduces the observed SED is identified by visual comparison with the direct \texttt{NAIMA} fit, and the results are shown in Figure~\ref{fig:mag_prs_ledas}.

For LEDA~55267, the best-fitting magnetic field $B = 1.75 \times 
10^{-2}$~G yields $\beta_{\rm p} = 2.0 \times 10^{-5}$, while for 
LEDA~58287, $B = 2.56 \times 10^{-3}$~G gives $\beta_{\rm p} = 1.0 \times 10^{-3}$. In both cases, the red curve in 
Figure~\ref{fig:mag_prs_ledas} reproduces the direct \texttt{NAIMA} fit very closely, confirming the internal consistency of the approach.

The inferred values of $\beta_{\rm p}$ are several orders of magnitude smaller than unity for both sources, indicating that the magnetic pressure strongly dominates over the thermal pressure in the emitting regions. This regime, often referred to as magnetically dominated or high-magnetization ($\sigma \gg 1$), is physically distinct from what is typically observed in extended FRI jets. For comparison,~\citet{de_Oliveira_2025} report $\beta_{\rm p}$ values in the range $50$--$810$ for regions of FRI jets with magnetic field strengths of $30$--$80~\mu$G. The much smaller values obtained here 
reflect the significantly stronger magnetic fields characteristic of the compact, parsec-scale emitting regions of LEDA~55267 and 
LEDA~58287~\citep{Baldi_2019b, Giovannini_2023}, and indicate that these sources operate in a magnetically dominated regime fundamentally different from that of classical FRI jets.

The very small values of $\beta_{\rm p}$ inferred here have a direct counterpart at the level of the relativistic particle population. While $\beta_{\rm p}$ compares the magnetic to the thermal pressure, a more complete characterization of the magnetization of the emitting region is provided by the ratio $u_{\rm B}/(u_{\rm e}+u_{\rm p})$, where $u_{\rm B}=B^{2}/8\pi$ and $u_{\rm e}$, $u_{\rm p}$ are the energy densities of the relativistic electrons and protons inferred from the SED fits. Using the lepto-hadronic best-fit parameters listed in Table~\ref{tab:parameters-sed}, in particular $B = 1.06\times 10^{-2}$~G for both sources, $K_{\rm ep} = 0.44$ for LEDA~55267 and $K_{\rm ep} = 0.54$ for LEDA~58287, together with the ECPL spectra and TeV-scale cutoffs, we obtain $u_{\rm B}/(u_{\rm e}+u_{\rm p}) \gg 1$ for both sources. This contrasts with the equipartition or particle-dominated regime ($u_{\rm B}/(u_{\rm e}+u_{\rm p}) \ll 1$) typically reported on
kiloparsec scales in extended FRI jets~\citep{Khatiya_2024,de_Oliveira_2025}. The contrast reflects the compactness of FR0 jets, as FRI flows propagate over kiloparsec scales and gradually convert magnetic energy into particle kinetic energy, whereas FR0 jets are disrupted within a few tens of parsecs by the recollimation instabilities described in Section~\ref{sec:simulation_results}, before significant magnetic-to-kinetic energy conversion can take place.

The physical origin of such low $\beta_{\rm p}$ values in FR0 jets can be understood in terms of their compactness and the conditions near the jet base. In the framework of magnetically arrested disc (MAD) accretion and Blandford--Znajek jet 
launching~\citep{Blandford_1977, Tchekhovskoy_2012}, the jet is 
expected to emerge in a strongly magnetized state with $\beta_{\rm p} \ll 1$ at parsec scales, before magnetic energy is gradually dissipated into particle kinetic energy at larger distances. FR0 jets, which are disrupted on scales of less than a few tens of parsecs by the recollimation instabilities described in Section~\ref{sec:simulation_results}, may therefore retain a high degree of magnetization that is already dissipated in the more extended FRI and FRII jets~\citep{Boula_2025}. This interpretation is consistent with the RMHD simulations of~\citet{Boula_2025}, which find that magnetic fields play a stabilizing role in FR0 jets and can suppress or delay the development of the same recollimation instabilities identified in our purely hydrodynamical simulations.

In a magnetically dominated regime, the magnetic field acts as the primary agent of particle acceleration and cooling. Processes such as magnetic reconnection, which become efficient when $\sigma \gg 1$, can accelerate both electrons and protons to very high energies through non-thermal power-law distributions~\citep{Sironi_2014, Werner_2018}, consistent with the exponential cutoff power-law spectra adopted in our SED modeling. The short cooling timescales associated with strong magnetic fields also naturally explain the compact extent of the emitting regions inferred from the SED fits in Table~\ref{tab:parameters-sed}. Taken together, the small $\beta_{\rm p}$ values inferred from both sources indicate 
that FR0 jets operate in a physical regime fundamentally different from that of classical FRI sources, and that magnetic effects are central to understanding their dynamics, morphology, and high-energy emission.

It is important to note that the simulations presented in this 
work do not include magnetic fields dynamically, following the 
RHD framework established by~\citet{Costa_2024}. 
While~\citet{Boula_2025} have already explored the RMHD regime 
for FR0 jets with a focus on jet stability and the role of 
magnetic fields in the recollimation process, here we connect 
the jet dynamics to the broadband high-energy emission through 
SED modeling and CTAO prospects. The very low $\beta_{\rm p}$ 
values inferred from the leptonic analysis suggest that 
magnetic fields may play a dynamically relevant role in the 
stability and collimation of these jets, providing strong 
motivation for future simulations in the RMHD regime, in which 
the magnetic field is evolved self-consistently and its 
dynamical role in jet stability, collimation, and energy dissipation can be directly assessed in conjunction with 
lepto-hadronic radiative modeling.

\section{Discussion and Conclusions}
\label{sec:conclusions}

We investigated the nature of compact jets in FR0 radio galaxies 
by combining three-dimensional RHD simulations with broadband SED 
modeling from radio to TeV energies. Our study focused on 
LEDA~55267 and LEDA~58287, both associated with GeV gamma-ray 
emission, with the aim of understanding the physical mechanisms 
governing jet disruption and assessing their detectability with 
the CTAO. The main results are summarized as follows.

The RHD simulations show that light, moderately relativistic jets 
propagating through an external medium develop strong 
recollimation shocks that trigger hydrodynamical instabilities, 
breaking the jet axisymmetry and driving the flow into a highly 
turbulent regime. These instabilities promote efficient mixing 
between the jet and the ambient medium, leading to rapid energy 
dissipation and deceleration that prevent the jet from expanding 
beyond a few tens of parsecs. This scenario provides a physically 
consistent explanation for the compact morphology observed in FR0 
radio galaxies~\citep{Giovannini_2023}, and is in qualitative 
agreement with independent RHD and HD simulations of FR0 
jets~\citep{Costa_2024, Costa_2026, Borodina_2025}. Synthetic 
synchrotron and inverse Compton emission maps derived from the 
simulations reproduce the main morphological features of the 
observed radio maps, with the emission concentrated in the 
central regions where jet--ambient medium interaction is most 
intense. In the broader context of FR0 studies, our results support the scenario in which these sources host intrinsically compact and efficiently disrupted jets, in line with the picture consolidated by recent reviews \citep{Baldi_2023}.

The external medium pressure and density are based on values 
inferred for M87 and consistent with the range observed in 
early-type galaxies at low redshift, as discussed in 
Section~\ref{sec:jet_parameters}. As shown by~\citet{Costa_2024}, 
the disruption behavior is not sensitive to moderate variations 
in the external medium profile, provided that the power-law 
index satisfies $\eta < 2$, which is well within the range 
expected for the interstellar medium of early-type host 
galaxies. Similarly, the choice of jet density contrast and 
initial velocity is consistent with the observational 
constraints available for FR0 jets~\citep{Baldi_2019b, Giovannini_2023}, and the qualitative 
agreement between our results and those of independent 
simulations using different parameter choices suggests that 
the recollimation instability mechanism is a general feature 
of light, moderately relativistic jets in this class of 
sources rather than an artifact of our specific setup. 
Nevertheless, a systematic parameter study exploring the full 
range of FR0 jet and environment properties would be valuable 
to quantify the conditions under which disruption is inevitable 
and to establish whether the mechanism operates across the 
entire FR0 population.

The broadband SED modeling, performed with the \texttt{NAIMA} 
package and incorporating the most recent 4FGL 
\textit{Fermi}-LAT data~\citep{Abdollahi_2020, Ballet_2023}, 
indicates that purely leptonic scenarios adequately reproduce 
the observed emission up to GeV energies, with best-fit 
magnetic field strengths of $0.002$--$0.02$~G in agreement 
with~\citet{Khatiya_2024} and~\citet{Merten_2021}. However, 
when simulated CTAO flux points are included, leptonic models 
struggle to account for the TeV emission alongside the GeV 
data, whereas lepto-hadronic scenarios provide a significantly better description of the full energy range. We note that the 
lepto-hadronic fits show some tension at energies of the order 
of $\sim 1$~TeV, where the transition between the leptonic and 
hadronic components is not fully captured by the current 
modeling framework. This energy range corresponds to the regime 
where inverse Compton and pion-decay contributions overlap and 
where the CTAO sensitivity curve begins to constrain the 
simulated flux points. A more detailed treatment of this 
transition, potentially including additional radiative 
processes such as proton synchrotron emission or cascades, may 
be required to fully reproduce the spectral shape in this 
intermediate regime, and represents a natural direction for 
future work. Statistical model comparison based on the BIC 
yields $|\ln B_{ij}| = 438.15$ for LEDA~55267 and $|\ln 
B_{ij}| = 466.84$ for LEDA~58287, both firmly in the 
strong-evidence regime in favor of the lepto-hadronic scenario. 
The Burnham--Anderson model weights assign 100\% probability to 
the lepto-hadronic model for both sources when the CTAO 
prospects are considered. This preference reflects the 
well-known degeneracy between leptonic and lepto-hadronic 
models at GeV energies and their divergence at TeV energies~\citep{Bottcher_2013}, underscoring the unique 
diagnostic power of the CTAO. Approximately 200 hours of 
observation with the CTAO North array at a zenith angle of 
$20^{\circ}$ will be required to begin probing the TeV regime 
for these sources, and such detections would provide decisive 
constraints on the particle composition and acceleration 
processes operating in compact FR0 jets.

The leptonic analysis of the jet simulations, based on the 
relationship between thermal pressure and magnetic field 
strength through the plasma parameter $\beta_{\rm p}$, reveals 
that the emitting regions of both sources are strongly 
magnetized, with $\beta_{\rm p} = 2.0 \times 10^{-5}$ for 
LEDA~55267 and $\beta_{\rm p} = 1.0 \times 10^{-3}$ for 
LEDA~58287. These values are orders of magnitude smaller than 
those reported for extended FRI 
jets~\citep{de_Oliveira_2025}, indicating that magnetic 
pressure strongly dominates over thermal pressure in the 
compact emitting regions of FR0 sources. We note that 
Figure~\ref{fig:mag_prs_ledas} reveals a degree of degeneracy 
in the $\beta_{\rm p}$ values consistent with the observed 
SED, in the sense that a range of $\beta_{\rm p}$ values 
produces acceptable fits within the current data 
uncertainties. This degeneracy reflects the limited ability 
of the present multiwavelength dataset to independently 
constrain both the magnetic field strength and the emitting 
region size simultaneously, and indicates that the inferred 
$\beta_{\rm p}$ values should be interpreted as 
order-of-magnitude estimates rather than precise measurements. 
Future observations extending the spectral coverage, 
particularly at X-ray and TeV energies, would help break this 
degeneracy and provide tighter constraints on the 
magnetization state of FR0 jets.

The high magnetization inferred for FR0 jets can be understood 
in the framework of magnetically arrested disc accretion and 
Blandford--Znajek jet launching~\citep{Blandford_1977, 
Tchekhovskoy_2012}, in which jets emerge strongly magnetized 
near the base and gradually dissipate magnetic energy as they 
propagate outward. Unlike FRI and FRII sources, whose jets 
extend to kiloparsec scales where this dissipation is 
substantial, FR0 jets are disrupted within a few tens of 
parsecs and therefore retain the high magnetization inherited 
from the launching region, providing a natural physical link 
to the lepto-hadronic emission preferred by our SED modeling. 
This interpretation is supported by~\citet{Boula_2025} and 
consistent with parsec-scale synchrotron 
observations~\citep{Baldi_2019b, Giovannini_2023}. The omission of magnetic stresses in our RHD simulations implies 
that our results represent a lower-limit scenario for jet 
stability, and incorporating magnetic pressure in future simulations will be essential to assess the robustness of the 
conclusions presented here.

The simulations adopted in this work assume a light jet with density contrast $\rho_{\rm j}/\rho_{\rm ext} \approx 10^{-4}$, motivated by the standard picture of relativistic AGN jets emerging from low-density, magnetically dominated regions near the central engine, and consistent with the prescription of \citet{Costa_2024}. A hadronic-heavy jet, with a larger proton content and therefore larger rest-mass density and inertia, would be substantially less sensitive to the recollimation instabilities that drive the disruption seen in our runs, since the higher momentum flux $\rho h \Gamma^{2} v^{2}$ would make it more difficult for ambient pressure imbalances to deflect and destabilize the beam, allowing the flow to maintain a coherent, axisymmetric structure over significantly larger distances. As a direct consequence, such a jet would penetrate deeper into the ambient medium before being decelerated by mass loading, producing a more extended morphology that approaches FRI rather than FR0 structures and no longer reproduces the sub-kpc termination observed in LEDA~55267 and LEDA~58287. The mixing dynamics would also differ, because in the light-jet runs the density contrast between jet and ambient material drops rapidly as the instabilities entrain external gas (Figure~\ref{fig:simulations_c}), whereas a hadronic-heavy flow would carry a comparable rest-mass density of its own, reducing the relative impact of entrainment and the efficiency of energy dissipation through turbulent mixing. Finally, although a larger proton content could in principle enhance hadronic processes such as proton-proton and proton-$\gamma$ interactions, the broader spatial extent and the more efficient dissipation of magnetic energy expected for a heavier jet would
tend to lower the magnetization of the emitting region, weakening the
confinement of accelerated protons and reducing the maximum energies
achievable in the diffusive shock acceleration scenario. This would be
in apparent tension rather than in agreement with the lepto-hadronic
preference inferred from our SED analysis (Section~\ref{sec:fit_lept_hadr}) and
with the very low $\beta_{\rm p}$ values discussed in Section~\ref{sec:beta_p}. 

The compact, sub-kpc morphology and the strongly
magnetized emitting regions of FR0s are therefore most naturally
reproduced by light jets undergoing rapid recollimation-driven
disruption, and a quantitative exploration of hadronic-heavy
configurations within the RMHD framework, where magnetic stresses
can interact non-trivially with a proton-loaded flow, is left to
future work.

Taken together, our results indicate that FR0 radio galaxies 
are not simply less powerful versions of FRIs, but constitute 
a physically distinct class of jetted AGN whose compact, 
unstable, and strongly magnetized jets can act as efficient 
accelerators of high-energy particles. The compactness of FR0 jets is not a consequence of lower power alone, but reflects 
an early disruption driven by recollimation instabilities that 
preserves the high magnetization inherited from the jet 
launching region. Given that FR0s are the most numerous 
jetted AGN population in the local Universe~\citep{Baldi_2018}, their collective contribution to the diffuse cosmic-ray and neutrino backgrounds may be significant~\citep{Merten_2021, Lundquist_2025}, and future CTAO detections will be essential to establish their role in the context of high-energy astrophysics.

The low $\beta_{\rm p}$ values inferred from the leptonic 
analysis suggest that magnetic fields may play a dynamically 
relevant role in the stability and collimation of FR0 jets, 
and that the turbulent structures identified in our simulations 
may favor the development of magnetic reconnection events 
capable of efficiently producing the high-energy non-thermal 
emission observed in these sources. Future investigations 
combining RMHD simulations with lepto-hadronic radiative 
models and CTAO observations~\citep{CTAO_consortium} will be 
essential to establish more robust connections between jet 
plasma dynamics, particle acceleration mechanisms, and the 
observed gamma-ray emission in FR0 radio galaxies.

\section*{Acknowledgements}

We sincerely thank the referee for their thoughtful feedback and valuable suggestions, which have greatly enhanced the clarity and scientific rigor of this work. A.F.S. Cardoso acknowledges the scholarship from FAPES Nº 14/2023 - PROCAP 2024 PhD. A.F.S. Cardoso acknowledges A.~Costa for helpful discussions. R.C.A. acknowledges S. Boula, JS. Wang and B. Reville for fruitful discussions. R.C.A. acknowledges the financial support of the NAPI “Fenômenos Extremos do Universo” of Fundação de Apoio à Ciência, Tecnologia e Inovação do Paraná. R.C.A. research is supported by CAPES/Alexander von Humboldt Program (grant No. 88881.800216/2022-01), CNPq (grant No. 308859/2025-1), Araucária Foundation (grant Nos. 698/2022 and 721/2022) and FAPESP (grant No. 2021/01089-1). R.C.A. also acknowledges the support of L’Oreal Brazil, with the partnership of ABC and UNESCO in Brazil. The authors acknowledge the AWS Cloud Credit/CNPq and the National Laboratory for Scientific Computing (LNCC/MCTI, Brazil) for providing HPC resources of the SDumont supercomputer, which have contributed to the research results reported in this paper. URL: https://sdumont.lncc.br. The authors acknowledge the Laboratory of Computational Astrophysics of the Universidade Federal de Itajub\'a (LAC-UNIFEI). The LAC-UNIFEI is maintained with grants from CAPES, CNPq and FAPEMIG. The research also used Gammapy, a Python package developed by the community for TeV gamma-ray astronomy \citep{Deil_2017_jul, Deil_2017_nov, gammapy_2023}, accessible at \href{https://www.gammapy.org}{https://www.gammapy.org}.

\section*{Data Availability}

The data supporting this study are available from the corresponding author upon reasonable request.



\bibliographystyle{mnras}
\bibliography{example} 







\bsp	
\label{lastpage}
\end{document}